\documentclass[twocolumn]{aastex631}

\usepackage[normalem]{ulem}

\begin{document}

\title{Seasonal thaws under mid-to-low pressure atmospheres on Early Mars}

\author[0000-0002-7744-5804]{Paolo Simonetti}

\affil{INAF - Trieste Astronomical Observatory,
Via G. B. Tiepolo 11,
34143 Trieste, Italy}

\author[0000-0001-7604-8332]{Giovanni Vladilo}
\affil{INAF - Trieste Astronomical Observatory,
Via G. B. Tiepolo 11,
34143 Trieste, Italy}

\author[0000-0002-8068-7695]{Stavro L. Ivanovski}
\affil{INAF - Trieste Astronomical Observatory,
Via G. B. Tiepolo 11,
34143 Trieste, Italy}

\author[0000-0002-7571-5217]{Laura Silva}
\affil{INAF - Trieste Astronomical Observatory,
Via G. B. Tiepolo 11,
34143 Trieste, Italy}
\affil{IFPU - Institute for Fundamental Physics of the Universe, Via Beirut 2, 34014 Trieste, Italy}

\author[0000-0002-6068-8682]{Lorenzo Biasiotti}
\affil{INAF - Trieste Astronomical Observatory,
Via G. B. Tiepolo 11,
34143 Trieste, Italy}
\affil{University of Trieste - Dep. of Physics, 
Via G. B. Tiepolo 11, 
34143 Trieste, Italy}

\author[0000-0001-9442-2754]{Michele Maris}
\affil{INAF - Trieste Astronomical Observatory,
Via G. B. Tiepolo 11,
34143 Trieste, Italy}

\author{Giuseppe Murante}
\affil{INAF - Trieste Astronomical Observatory,
Via G. B. Tiepolo 11,
34143 Trieste, Italy}
\affil{IFPU - Institute for Fundamental Physics of the Universe, Via Beirut 2, 34014 Trieste, Italy}
\affil{CNR - Institute of Geosciences and Georesources, 
Via Giuseppe Moruzzi, 1,
56127 Pisa, Italy}

\author{Erica Bisesi}
\affil{INAF - Trieste Astronomical Observatory,
Via G. B. Tiepolo 11,
34143 Trieste, Italy}
\affil{IFPU - Institute for Fundamental Physics of the Universe, Via Beirut 2, 34014 Trieste, Italy}
\affil{CNR - Institute of Geosciences and Georesources, 
Via Giuseppe Moruzzi, 1,
56127 Pisa, Italy}

\author{Sergio Monai}
\affil{INAF - Trieste Astronomical Observatory,
Via G. B. Tiepolo 11,
34143 Trieste, Italy}



\begin{abstract}

Despite decades of scientific research on the subject, the climate of the first 1.5 Gyr of Mars history has not been fully understood yet. Especially challenging is the need to reconcile the presence of liquid water for extended periods of time on the martian surface with the comparatively low insolation received by the planet, a problem which is known as the Faint Young Sun (FYS) Paradox. In this paper we use ESTM, a latitudinal energy balance model with enhanced prescriptions for meridional heat diffusion, and the radiative transfer code EOS to investigate how seasonal variations of temperature can give rise to local conditions which are conductive to liquid water runoffs. We include the effects of the martian dichotomy, a northern ocean with either 150 or 550 m of Global Equivalent Layer (GEL) and simplified CO$_2$ or H$_2$O clouds. We find that 1.3-to-2.0 bar CO$_2$-dominated atmospheres can produce seasonal thaws due to inefficient heat redistribution, provided that the eccentricity and the obliquity of the planet are sufficiently different from zero. We also studied the impact of different values for the argument of perihelion. When local favorable conditions exist, they nearly always persist for $>15\%$ of the martian year. These results are obtained without the need for additional greenhouse gases (e.g. H$_2$, CH$_4$) or transient heat-injecting phenomena (e.g. asteroid impacts, volcanic eruptions). Moderate amounts (0.1 to 1\%) of CH$_4$ significantly widens the parameter space region in which seasonal thaws are possible.


\end{abstract}


\section{Introduction} \label{sec:intro}

The early history of the planet Mars is divided into three periods called Noachian, Hesperian and Amazonian \citep[see e.g.][]{hartmann01}, each of them linked to a set of geographical regions grouped on the basis of density, size and superposition of craters. A fourth period, called pre-Noachian, precedes the most ancient terrains identified so far. The chronological intervals usually associated to these periods are 4.5 - 4.1 (pre-Noachian), 4.1 - 3.7 (Noachian), 3.7 - 3.0 Gyr ago (Hesperian), with the Amazonian continuing up to present day \citep[see e.g.][]{carr10}.

A wide range of geomorphological and chemical evidence suggests that, at some point in the past, liquid water was present on the surface of Mars for extended (at least $10^5$ - $10^7$ yr) periods of time \citep{hoke11,balme20} and in sufficient quantities to sustain a vigorous hydrological cycle \citep[e.g.][]{mangold04}. Since most of the dendritic valley networks are mainly associated with Noachian and, to a lesser extent, Hesperian terrain, the humid period of the martian history is identified with the Noachian and Hesperian ages ($\sim$4.1 - 3.0 Gyr ago). However, at that epoch, the power output of the Sun was most probably lower than today by $20-25\%$ \citep{gough81}, making it difficult to explain how Mars could have hosted climate conditions sufficiently warm to allow for such an hydrological cycle to be present. This longstanding problem is known as the Faint Young Sun (FYS) paradox.

The study of planetary climates is performed using a variety of different climate models. These models are usually categorized depending on their complexity, which scales to the number of dimensions assigned to the planet and its atmosphere. The simplest ones are called Energy Balance Models (EBMs) and can be either single-column or seasonal-latitudinal. Single-column EBMs derive the climatology of the entire planet from a single atmospheric column in radiative-convective equilibrium (RCE) with fixed surface albedo and thus focus on the radiative transfer (RT) properties of a given atmospheric composition \citep{pierrehumbert10}. Seasonal-latitudinal EBMs add the latitudinal horizontal direction and, as such, are able to estimate the evolution over time of local conditions on a planet surface \citep{north81}. A simplified horizontal heat diffusion might or might not be present, but only the former class of models can be applied to planets with non-negligible atmospheres and take into consideration some simple climate feedbacks, like the ice-albedo feedback. While historically applied to the study of fast rotating planets, seasonal-latitudinal EBMs have been successfully employed to model also tidally-locked ones \cite[e.g.][]{kite11,haqqmisra22}. Recently, \cite{okuya19} proposed an EBM that also considers the longitudinal heat diffusion and is thus capable of producing full surface temperature maps.
At the upper end of the complexity spectrum there are the General Circulation Models (GCMs), which are coupled radiative transfer-fluidodynamical models that integrate the Navier-Stokes equations to reconstruct the full 3D motions of a planetary atmosphere \citep[e.g.][]{richardson07}. The required computational power is proportional to the model complexity, which makes GCMs unsuitable for parameter space explorations.


In the last decades, 
many possible solutions to the FYS paradox have been proposed, that can be generally 
classified in two large groups. 
The first one searches for an atmospheric composition and thickness capable to warm the surface up to the required temperature. Early calculations focused on the viability of a thick CO$_2$-H$_2$O atmosphere, but in single-column EBMs this is both insufficient to achieve the goal \citep{kasting91} and difficult to reconcile with observational data concerning the surface paleopressure of Mars \citep[][but see also \citeauthor{bultel19}, \citeyear{bultel19}]{lammer13,kite14,edwards15}. More recent investigations took into consideration the greenhouse effect produced by other chemical species such as SO$_2$/H$_2$S \citep{halevy07,johnson08,mischna13} or H$_2$/CH$_4$ \citep{ramirez14,wordsworth17,ramirez17b,haberle19,turbet20b}, satisfactorily solving the issues associated with CO$_2$-only atmospheres, but creating new ones related to the availability of these additional gases \citep[see e.g.][]{wordsworth16a}. The role of infrared backscattering from CO$_2$ clouds to heat the planet has also been investigated \citep{forget97}, but later calculations cast doubts on its effectiveness \citep{colaprete03,kitzmann13,forget13}, since a vigorous greenhouse effect requires a very specific and narrow distribution of cloud particle sizes. Similar conclusions were drawn concerning the radiative forcing of H$_2$O clouds \citep{wordsworth13,urata13,ramirez17a}. Variations of the martian surface albedo were also taken into consideration but discarded \citep{fairen12}. 
The second group of proposed possibilities focuses on the role of transient warmings as solutions to the FYS paradox, such as asteroid impacts \citep{segura02,segura08}, volcanic eruptions \citep{halevy14}, the destabilization of previously formed CH$_4$ clathrates due to chaotic obliquity changes \citep{kite17} or episodic injections of H$_2$ due to crustal alteration \citep{wordsworth21}. None of them, however, are free of issues. The post-impact rainfall pattern predicted by more modern models mismatches observations \citep{turbet20a}; photochemically driven formation of highly reflective sulfate aerosols from volcanic SO$_2$ and H$_2$S may have cooled, rather than warmed, the planet \citep{tian10,kerber15}; it is uncertain if CH$_4$ could have been produced and stored in sufficient quantities in the martian environment in the first place \citep{ramirez18}.

The two sets of solutions proposed for the FYS paradox give rise to two possible Early Mars climatologies, called ``warm-and-wet" \citep[e.g.][]{craddock02} and ``cold-and-icy" \citep[e.g.][]{squyres94}. The warm-and-wet scenario is associated with a hydrosphere which rests mostly in a relatively wide ocean covering a considerable portion of the northern lowlands (plus the Hellas and Argyre basins in the south), with minor and/or transient glaciers on the southern highlands. This scenario offers the most straightforward explanation for valley networks formation and is in accordance with the existence of a peak in the altitude distribution of fluvial delta-like formations corresponding to the paleo-shoreline of the northern ocean \citep{diachille10}. This scenario however, has been criticized because it does not seem to be able to produce the precipitation pattern required to form the observed valley networks in the right places \citep[see e.g.][]{wordsworth15}. Also, in order to prevent the locking of the hydrosphere in the southern highlands (which act as an efficient cold trap), an even higher global average temperature is required, which puts an extra constraint on the atmospheric chemical composition. 
The cold-and-icy scenario envisions instead a situation where most of the water reservoirs are locked in glaciers and episodic melting events provide the liquid water runoffs necessary to explain surface erosion. In the case of a large ($\gtrsim 200$ m) global equivalent layer (GEL) of water, significant basal melting can occur at the bottom of highland ice sheets, allowing for efficient fluvioglacial erosion \citep{fastook14}. However, the lack of an extensive glacial erosion in the regions where these ice sheets should have been existed is a major unsolved issue of this picture.

The duration and nature of the conditions allowing for liquid water on the martian surface have important astrobiological implications, too. The possible survival of martian life in subsurface environments up to the present day, which is occasionally invoked to explain seasonal variabilities in the martian atmospheric composition \citep{webster15}, hinges on the existence of a climate state conductive of the emergence of life in the first place. 
While the conditions for abiogenesis are still unknown, 
early Mars is believed to hold a widespread range of physical and chemical properties potentially suitable for an independent origin of life, with different possible niches depending on the prevalent climate state \citep[e.g.][]{clark21}.

%

Most of Early Mars modelings have been performed either via parameter space explorations with single-column EBMs, or by running very small and specific sets of cases with 3D GCMs. This is sound considering the respective computational costs of these models and yielded a vast amount of insights on the surface conditions during the Noachian and Hesperian periods. Efficient single-column EBMs have been coupled with a variety of other codes, modeling processes such as the atmospheric photochemistry \citep{batalha15}, the carbonate-silicate cycle \citep{batalha16} or ecological interactions \citep{sauterey22}. On the other hand, GCMs have been fundamental to cast light on the precipitation pattern and ice deposition \citep[e.g.][]{wordsworth15,kamada20,kamada21}. In this work, we make the case for the employment of seasonal-latitudinal EBMs coupled with a flexible RT code as the natural next step along the path inaugurated by single-column models by studying the effects of seasonal surface temperature variations, especially under the influence of specific choices regarding the orbital elements of Early Mars (eccentricity, obliquity and the argument of perihelion). In particular, we use EOS-ESTM \citep{biasiotti22} to find the maximum and minimum latitudinal band-averaged temperatures to see if a given set of atmospheric compositions (including CO$_2$, H$_2$O and CH$_4$) and orbital parameters are able to produce seasonal melting of surface ices without concurrently causing seasonal condensation of the atmosphere at surface. We also tested different hypotheses in terms of surface albedo and cloud radiative effects. Using seasonal-latitudinal EBMs, either non-diffusive \citep[e.g.][]{armstrong04,kite13} or diffusive \citep{ramirez20,hayworth20} is not entirely new, but it is still largely unexplored and promises to yield new results by taking the best of both worlds, i.e. retaining a high computational efficiency while allowing a better understanding of the conditions that might arise on the planetary surface.

The structure of this paper is as follows. In Section 2 we describe our climate (ESTM) and vertical radiative transfer (EOS) models, together with the specific choices we made to take into consideration the inhomogeneous water and altitude distribution of Mars. In Section 3 we show the results of the EOS-ESTM parameter space explorations. In Section 4 we discuss our results, in particular comparing them with some bulk observational evidences and in Section 5 we draw our conclusions.

\section{Model description} \label{sec:models}

\subsection{ESTM} \label{ss:estm}

\begin{table}[]
\centering
\begin{tabular}{lcc}
\hline
Parameter & Description & Value \\
\hline
$a$ & semi-major axis & 1.524 AU \\
$e$ & eccentricity & 0.0934 \\
$I$ & insolation & 439.5 W m$^{-2}$ \\
$\varepsilon$ & obliquity & 25$^\circ$ \\
$\omega$ & argument of perihelion & 286.5$^\circ$ \\
$P_{\text{rot}}$ & rotation period & 1.0275 d \\
$P_{\text{orb}}$ & orbital period & 686.96 d \\
$g$ & surface gravity acceleration & 3.72 m s$^{-1}$ \\
$R$ & mean radius & 3389 km \\
\hline
\end{tabular}
\caption{Orbital and planetary parameters adopted for the Early Mars in this work when not otherwise specified.}
\label{tab:orbital_params}
\end{table}

The Earth-like planet surface temperature model (ESTM) is a largely upgraded version of a latitudinal and seasonal Energy Balance Model (EBM). In common with classic EBMs \citep{north81}, the meridional transport is treated with a formalism of heat diffusion. In practice, in each latitude zone the energy balance is described by the equation
\begin{equation}
C  \frac{\partial T}{\partial t} - 
\frac{\partial}{\partial x}
\left[ D \, (1-x^2) \, \frac{ \partial T}{\partial x} \right]
+ I = S \, (1-A) ,
\label{eq:diffusion}
\end{equation}
where $T$ is the zonal temperature mediated over one rotation period and all terms are normalized per unit area. Most of the coefficients depend on both time, $t$, and latitude, $\varphi$, either directly or indirectly, through their dependence on $T$. 
The output of ESTM is $T$ as a function of $\varphi$ and the orbital phase.
The first term of the equation represents the seasonal evolution of the zonal heat storage; $C$ is the zonal heat capacity. The second term  represents the amount of heat per unit time leaving the zone along the meridional direction; $D$ is the ``diffusion term'' and $x=\sin \varphi$.
The term $I$ is the thermal radiation emitted by the zone, also called Outgoing Longwave Radiation (OLR). The right hand of the equation represents the zonal heating due to stellar photons; $S$ is the incoming stellar radiation 
and $A$ the albedo at the top of the atmosphere (TOA albedo).

At variance with classic EBMs, where $D$ is a constant, ESTM incorporates a physically-based description of $D$ \citep{vladilo15}. 
If we define $\Phi$ in such a way that $N = 2 \pi R^2 \Phi \cos \varphi$ is the net rate of energy transport across a circle of constant $\varphi$ in a planet of radius $R$ \citep[][]{pierrehumbert10}, the coefficient $D$ can be derived as a function of physical quantities from the analogy with the equation of heat diffusion, i.e. $\Phi \equiv -D (\partial T/\partial \varphi)$. By introducing an analytical description of the longitudinally averaged atmospheric transport validated with 3D climate models, $D$ is eventually calculated as a function of $R$, surface atmospheric pressure, $P$, surface gravity acceleration, $g$, angular rotation rate, $\Omega$, and other planetary properties \citep{vladilo15}. 

Another substantial improvement of ESTM with respect to classic EBMs concerns the description of the TOA albedo, $A$, and of the OLR, $I$. In classic EBMs, $A$ is a constant and $I$ a linear function of $T$. In ESTM, $A$ and $I$  are calculated by taking into account the atmospheric radiative transfer in the short and long wavelength ranges, respectively. In practice, this is achieved by interpolating on previously calculated lookup tables in which $A$ and $I$ for a specific atmospheric composition and pressure are stored as a function of the surface temperature $T$ and other relevant quantities. These calculations are performed for clear-sky atmospheres of given chemical composition and surface pressure, making use of the EOS code described below. Clouds are then incorporated in ESTM through a parameterization of their impact on the albedo and on the OLR. Starting from a realistic description of the surface albedo, the TOA albedo is calculated by taking into account the atmospheric transport in the short wavelength range and the simplified treatment of the cloud albedo. Both the cloud and the surface albedos are dependent on $\mu = \cos z$, where $z$ is the zenith angle of the incident light.

Apart for the parameters that we explicitly changed and that we discuss below, we adopted the same input values as in \cite{biasiotti22}.


\begin{figure*}
\plottwo{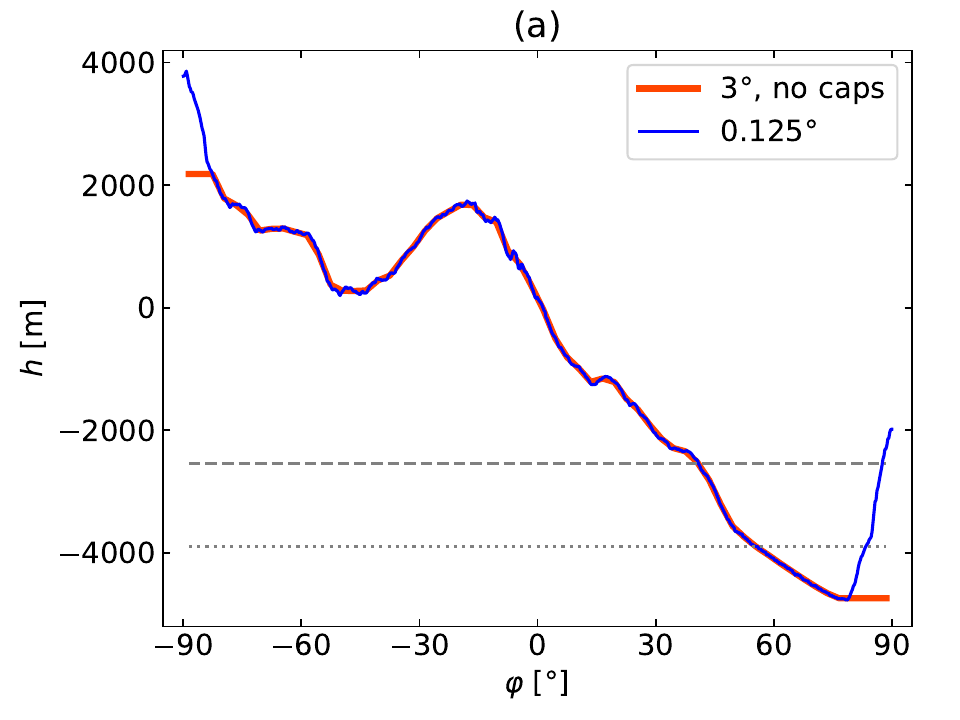}{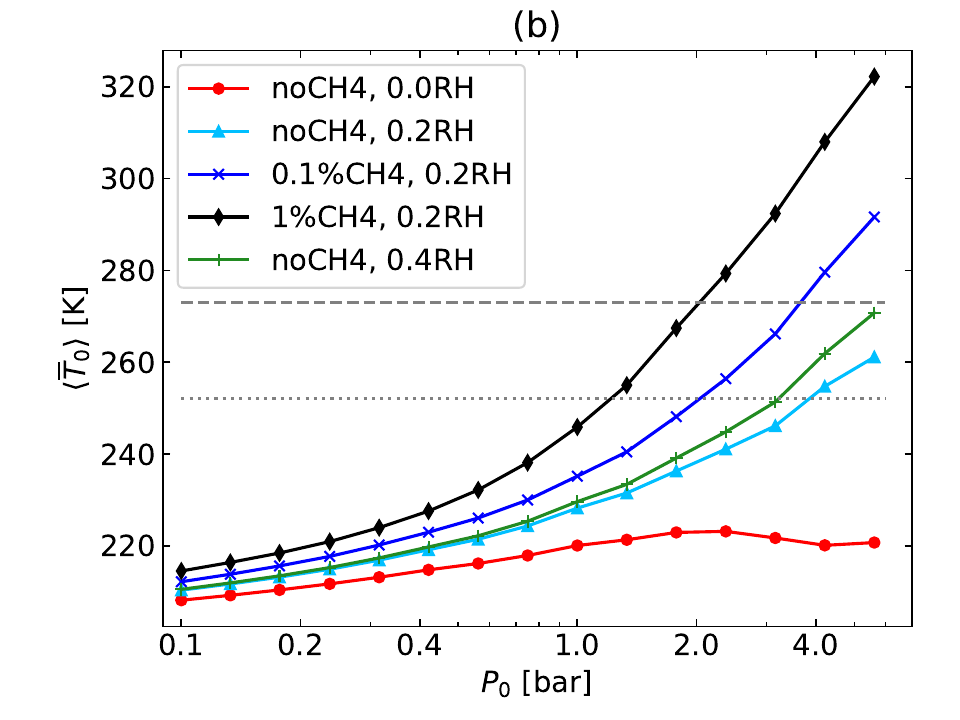}
\caption{Panel (a): band-averaged altitude of modern Mars surface with respect to the altimetric datum, calculated from the MOLA Mission Experiment Gridded Data Records (MEGDRs) map. Blue line: original resolution of 0.125° per band. Red line: ESTM resolution of 3° per band and modern ice caps removed. Gray dashed and dotted lines: isolines corresponding to a 550-m GEL ocean and a 150-m GEL ocean, respectively. Panel (b): The yearly global average planetary temperature at the altimetric datum $\langle\overline{T}_0\rangle$ for different values of the surface pressure. Red, cyan, blue, black and green lines refer respectively to the A0F0, B0G0, BLG0, BHG0 and C0G0 models described in Table \ref{tab:models}. Each marker represents an increase of 33\% of the pressure with respect to the preceding one. Grey dashed and dotted lines mark the 273 K and the 252 K level, respectively. \label{fig:altimetry_and_avgtemp}}
\end{figure*}

\subsubsection{Topography}\label{sss:topography}

Mars shows a marked topographical dichotomy between the northern and southern hemisphere. Approximately a third of the martian surface is composed of relatively young lowland plains, while the rest are highly cratered highlands. This geographical feature precedes the impact that formed the Hellas basin \citep[around 4.0 Gyr ago,][]{fassett11} and has been attributed either to plate tectonics \citep{sleep94}, to mantle convection \citep{zhong01,sramek12} or to a giant impact \citep{wilhelms84,andrews-hanna08,marinova08}.

The martian dichotomy deeply influences the climate of Mars, especially in presence of a thicker atmosphere. First, it impacts the global general circulation, changing how heat is horizontally redistributed on the surface. Second, it changes the actual surface temperature $T_s$ with respect to the temperature at a given reference altitude (specifically, the altimetric datum, see below) $T_0$ via the atmospheric lapse rate.

ESTM calculates the seasonal-latitudinal map of temperatures $T(t,x)$ at a single altitude across the entire planet. In this paper, we refer to these temperature values as $T_0$. However, since the martian dichotomy is mostly a latitudinal feature, we were able to include its effects by adjusting the ESTM output for the band-averaged lapse rate, thus deriving the actual surface temperature, $T_s$.
In practice, we followed the steps described below. First, we took the altimetric map with a 0.125° horizontal resolution generated by the Mars Orbiter Laser Altimeter (MOLA) experiment \citep{smith03} and calculated the band averaged altitude $h$ of the surface. We then binned the resultant $h$ at a 3° resolution in order to use it in ESTM and removed the ice caps, defined as the excess altitude above 77° and below -83° in latitude. The band averaged altitude profiles that we calculated are shown in Fig.~\ref{fig:altimetry_and_avgtemp}, panel (a). As in the original MOLA maps, the zero-altitude point (also called altimetric datum) is the average Mars radius at the equator. Then, we calculated the lapse rate for each seasonal and latitudinal point of our runs. The lapse rate of the martian atmosphere changes with temperature both in the dry case, where it can be ascribed to changes in the specific heat, and in the moist case, where it is caused mainly by different amounts of water vapor in the column. In particular, in this work we tested models with relative humidities ($RH$) equal to 20\% and 40\% (see Section \ref{sss:atmospheres}). The equations we used to calculate the lapse rate are described in Appendix A. 
We find the lapse rate to be limited between 4.2 and 5.2 K km$^{-1}$ across all models. This translates to adjustments of the surface temperature that are $\le$10 K across most of the martian surface, reaching $\sim$25 K in the most extreme cases. Finally, we applied these temperature adjustments, obtaining $T_s$ maps in place of the $T_0$ maps produced by ESTM. The results presented in Section \ref{sec:results} and discussed in Section \ref{sec:discussion} are based on the post-processed $T_s$ maps.

A third effect of the martian altimetry on the surface temperature is caused by the amount of atmosphere above different latitudinal bands. This changes the radiative exchanges between the surface and the outer space, decreasing the OLR and increasing the TOA albedo locally. With respect to the pressure at the altimetric datum $P_0$, the surface pressure $P_s$ at the highest latitudinal band in our topographical model (the South Pole, at 2180 m) is 15\% lower, while its value at the lowest point (the North Pole, at -4740 m) is 40\% higher. 
While this radiative effect can not be directly included in ESTM, we estimated its impact by considering how the average temperature of our models is influenced by $P_0$, as done e.g. by \cite{forget13}. The result is shown in Fig.~\ref{fig:altimetry_and_avgtemp}, panel (b), for the five atmospheric compositions and cloud models considered in this work (see Table \ref{tab:models}). The spacing in pressure is logarithmic and each pressure value represents an increase of 33\% with respect to the preceding one. As such, the temperature difference between a marker in the plot and its two nearest neighbours can be taken as an approximation of the net radiative effect that is expected to occur at different locations on the martian surface. This difference is largest in the high methane models (BHG0 and BHGH), at the second highest tested pressure (4.22 bar) and amounts to $\sim$13 K, while it is smallest in the dry, CO$_2$-only model at around 2 bar, where it amounts to less than 0.2 K. At low ($<$0.6 bar) pressures, where correctly accounting for the surface temperature is important to assess whether the atmosphere collapses or not, the difference is of the order of $\sim$2 K, which is four to ten times smaller than the effect of the lapse rate. The actual difference would be smaller due to the redistribution of heat, and this effect is stronger in denser atmospheres.

The columnar mass of atmosphere is influenced also by the oblateness of Mars and the latitudinal variation of $g$ due to planetary rotation. While these minor effects are not accounted for in ESTM, we can try to estimate them. The tangential velocity at the martian equator is $\sim$240 m s$^{-1}$, which decreases the perceived value of $g$ by 0.017 m s$^{-2}$. With respect to the average $g$ of Mars, this amounts to a reduction of $\sim$0.45\%. The difference between the polar and equatorial radii \citep[3376 km and 3396 km, respectively,][]{seidelmann07}, causes $g$ to decrease by $\sim$0.4\%, thus the combined effect amounts to $\sim$0.9\%. This variation in the columnar mass of the atmosphere is expected to cause, at maximum, a temperature increase in the equatorial regions of 0.35 K with respect to our calculations. In most of the tested cases and at higher latitudes, however, the effect is much smaller.

\subsubsection{Ocean fraction}\label{sss:oceans}

A number of independent clues seems to suggest the existence of a water ocean on the northern hemisphere of Mars at some point in the past, and most probably during the Noachian and Hesperian periods ($\sim$4.1-3.0 Gyr ago). Just a few examples of them are the similar height of ancient river outlets deposits with respect to the datum \citep{diachille10}, the sedimetary-like dielectric properties of the northern terrain \citep{mouginot12} and the shape of certain northern craters suggestive of an impact in shallow waters \citep{costard19}. On the other hand, the ability to support a large, open ocean under the FYS conditions is central in the debate between supporters of a warm-and-wet and a cold-and-icy Early Mars \citep[see e.g.][]{wordsworth16a}.

The global equivalent layer (GEL) of water present on  Early Mars is still a matter of debate, with estimates ranging from 100 m to 1500 m \citep[e.g.][]{scheller21}. The GEL also underwent evolution over time due to sequestration in the crust \citep{scheller21} and/or photodissociation and atmospheric escape \citep{webster13,villanueva15}, becoming smaller and smaller during the Noachian and Hesperian ages until it reached the current estimated value of 10-20 m \citep{plaut07}. 


In this work we tested three different scenarios in terms of surface water: (i) no oceans, (ii) oceans produced by a 150-m GEL and (iii) oceans produced by a 550-m GEL. The first scenario considers the case of a planet where water covers a negligible fraction of the surface. This includes the case in which most of the water is locked in glaciers or subsurface lakes \citep[]{orosei18}. The second scenario considers the long-term water level identified by \cite{carr03}, \cite{ivanov17} and \cite{salese19} and adopted also in other studies \citep[e.g.][]{schmidt22}, and explores the lower end of the estimated Early Mars water GEL. Since most of the evidences on which it rests are linked to Hesperian features, this can be also related more specifically to the conditions of the Hesperian Mars. The third and last scenario adopts instead the estimates of \cite{diachille10} and is somewhat in the middle of the Early Mars water GEL interval. This is more relevant for the study of a late Noachian Mars.

\begin{figure*}
\plottwo{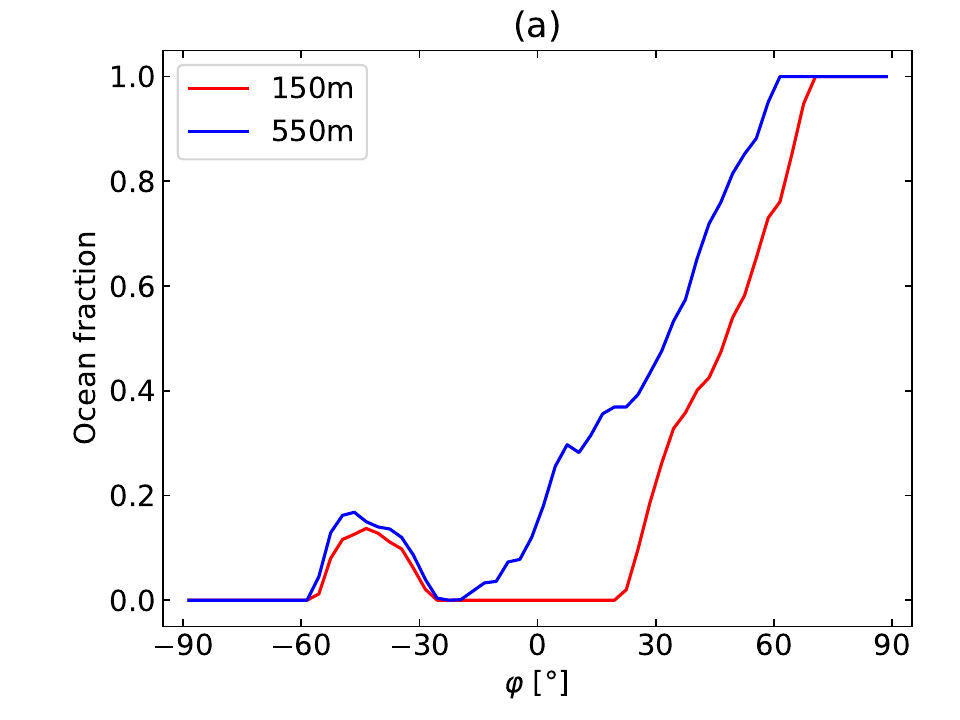}{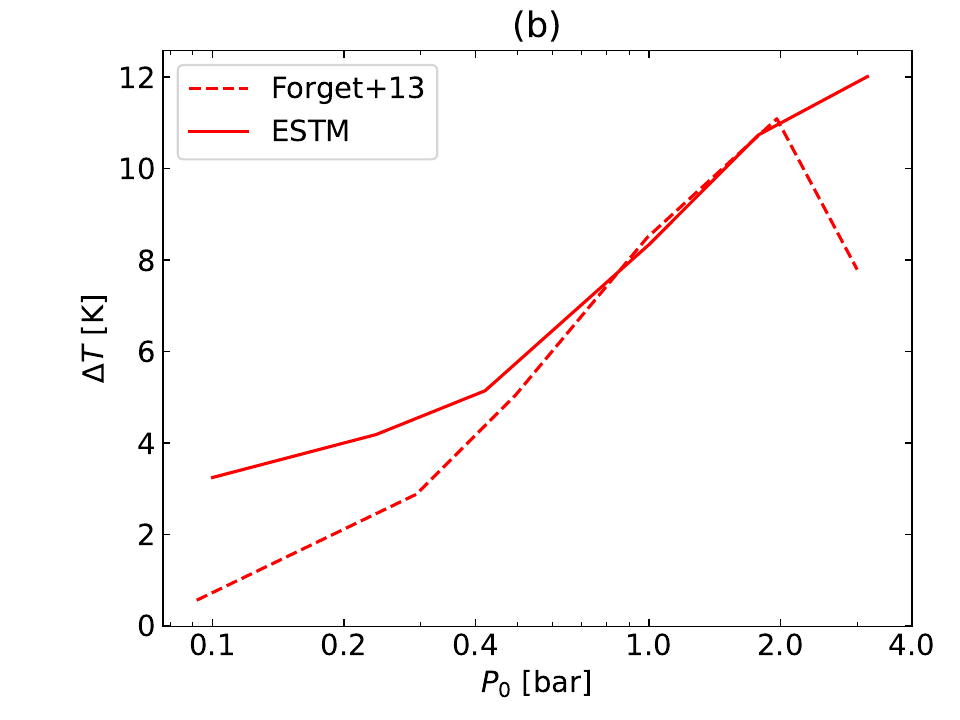}
\caption{Panel (a): band averaged ocean fraction based on the position of the martian paleo-shoreline in the models with a 150-m GEL \citep[late Hesperian ocean, as modeled by][]{schmidt22} or in the model with a 550-m GEL \citep[Noachian ocean, as in][]{diachille10}. Panel (b): CO$_2$ cloud effects on the average planetary temperature with respect to the clear-sky case in the \citet[dashed line]{forget13} and in the A0F0 model tested in this work (see Tab~\ref{tab:models}). \label{fig:oceans_and_clouds}}
\end{figure*}

ESTM allows us to specify the ocean fraction for each latitudinal band. This is usually not useful when dealing with exoplanets \citep[see e.g.][in which the ocean fraction is the same at each latitude]{silva17}. However, the Noachian and Hesperian martian topography is reasonably constrained, as it is its impact on the past distribution of water on the planet.
Thus, we derived the latitudinal distribution of the martian paleo-ocean from the same topographical map employed for estimating the latitudinal altitude profile and used it in our climate model. Among different paleo-shorelines identified in literature, we adopted the one at the -3900 m height \citep{carr03} in scenario (ii), and the one at the -2540 m height \citep{diachille10} in scenario (iii). At this point we averaged over all longitudes to obtain the latitudinal ocean fraction distribution. The curves obtained for scenarios (ii) and (iii) are shown in Fig.~\ref{fig:oceans_and_clouds}, panel (a), and correspond to the global ocean coverage of 0.16 and 0.30, respectively.

\subsubsection{Surface ices}\label{sss:ices}

The martian topography has an important impact also on the distribution of surface ice. First of all, the effect of the lapse rate may prevent the freezing of the low-lying water bodies even if the datum temperature is below the freezing point of water. Second, the southern highlands and the Tharsis bulge tend to act as cold traps, blocking most of the martian water reservoir and decreasing the planetary albedo \citep{wordsworth16a}.

ESTM lacks the ability to model the deposition and transportation of ice and instead evaluates the fractional surface covered by ice at a given latitude as a function of $T_s$ averaged over 12 months. 
In ESTM, the dependence of the ice coverage on temperature is given by a generalized logistic function with two different sets of parameter values to describe land and ocean ice coverage \citep[Eq.~4 in][]{biasiotti22}. These parameters have been tuned to reproduce Earth data. In general, this approach is adequate to describe Earth-like seasonal deposition of snow on lands and formation on ice on oceans at high latitudes. 

Since we decided to model Early Mars as an essentially dry planet, with a desert-like average humidity, we self-consistently expect only sporadic precipitations across the southern highlands. Thus, as far as lands are concerned, we choose to turn off the temperature-dependent algorithm of ice formation. To explore the impact of ice over land, we instead calculated the seasonal temperature variations for a range of average land albedos $\alpha_l$ (see Sect.~\ref{ss:surfalbedo_results}) up to values consistent with a permanent ice sheet of 50\% ($\alpha_l=0.41$).
%
%
On the other hand, as far as oceans are concerned, we kept the  temperature-dependent algorithm of  ice formation. The algorithm was fed by the surface temperature of the oceans  calculated with 
the procedure described in Sect.~\ref{sss:topography}. 
With respect to the temperature of the datum, we find that the surface ocean temperature is $16-20$ K higher in scenario (ii), and $10-13$ K higher scenario (iii), depending on the specific lapse rate obtained in a given run. Due to the band-averaged nature of our calculations, the actual ocean surface temperature is adjusted within $\sim$2 K at most for models using 150 m of water GEL and $\sim$1 K at most for the model using 550 m of water GEL.


As a minor note, we underline the fact that, since the original value of the turning point temperature of the logistic for the ice coverage fraction has been calibrated on Earth's oceans, it best describes the behaviour of water with the same average content of salt of Earth water, i.e. 3.5\% \citep{chester12}. This corresponds to a freezing temperature of 271.3 K.

\subsubsection{CO2 and H2O cloud modeling}\label{sss:clouds}

Cloud modeling is generally the strongest source of uncertainty when working with climate simulations in conditions that are different from those of the modern Earth atmosphere \citep[see e.g.][]{pierrehumbert10}. Clouds impact the climate of a planet both by increasing the albedo, which cools down the planet, and by reducing the OLR, which heats up the planet. These mechanisms act together and the net effect depends on a large amount of factors that are often difficult to determine, such as the height at which clouds typically form, the distribution of particle sizes inside the cloud and the total amount of condensate in the column \citep{ramirez17a}.



In ESTM clouds are parameterized via seven parameters linked to simple physical or geographical properties, which are summarized in Tab.~\ref{tab:clouds} and described below. The effect of clouds on the OLR is modeled via a cloud OLR forcing parameter (Cloud Radiative Effect, $CRE$), which can be fixed or made dependent on the average planetary temperature. The effect of clouds on the TOA albedo are modeled via three parameters: the cloud albedo, $a_{c}(\mu=0.5)$, the slope of the linear dependence between $\mu$ and the cloud albedo, $m_c$, and the cloud shortwave transmittance, $t$. The overall impact of clouds then scales linearly with the fractional cloud cover over each latitudinal band. In ESTM we specify different cloud coverage fractions for each of the three different planetary surfaces considered in the model: $f_{cw}$ for water, $f_{cl}$ for land and $f_{ci}$ for ice. Thus, the final cloud coverage at a given latitude is the weighted average between the three 
values. The functional dependence of the parameters described in this paragraph on the main radiative terms considered in ESTM are detailed in Section 2.5 of \cite{biasiotti22}.

\begin{table}[]
\centering
\begin{tabular}{lccc}
\hline
Input & CO$_2$ ice & H$_2$O Earth-like & ESTM \\
param. & clouds & clouds  &  calibration \\
\hline
$CRE$ & 15.0 & 26.1 & 5.2-26.1 \\
$a_{c}(0.5)$ & 0.28 & 0.44 & 0.44 \\
$m_c$ & -0.67 & -0.67 & -0.67 \\
$t$ & 0.95 & 0.85 & 0.85-0.95 \\
$f_{cw}$ & 0.5 & 0.5 & 0.72 \\
$f_{cl}$ & 0.5 & 0.5 & 0.55 \\
$f_{ci}$ & 0.5 & 0.5 & 0.56 \\
\hline
\end{tabular}
\caption{Parameters adopted to model CO$_2$-ice clouds (second column), Earth-like clouds (third column) and adopted in the ESTM reference paper \cite{biasiotti22} (fourth column). Values for $CRE$ and $t$ in the fourth column are expressed as an interval since they are temperature-dependent.}
\label{tab:clouds}
\end{table}

In this work, we tested two different models for the Early Mars clouds. The adopted values for the parameters
are reported in Table \ref{tab:clouds}. In order to relate more directly the input choices and results, we deactivated the dependence of $CRE$ and $t$ on $T_s$. The first model tries to capture the impact of CO$_2$-ice clouds. In practice, we tuned the parameters in ESTM as to obtain an overall average $T_s$ difference with respect to the clear-sky scenario that is comparable with the results of \cite{forget13}. In that paper, the thermal enhancement caused by clouds is explored for a range of surface pressure comprised between 0.1 and 6.5 bar, however the authors report that simulations with pressures above 3 bar are out-of-equilibrium because of atmospheric collapse. The comparison between our results and those of Forget and collaborators is shown in Fig.~\ref{fig:oceans_and_clouds}, panel (b). 

The second model is instead similar to that adopted in \cite{biasiotti22} to describe the modern Earth clouds. The values chosen for $CRE$ and $t$ are adequate for clouds that are present over non-icy surfaces (i.e., outside the Poles). \cite{biasiotti22} report that the exact choice of $t$ has only minor impacts on the final result.

At variance with \cite{biasiotti22} we fixed the fraction of clouds over all surfaces $f_c$ to the same value, in both models with CO$_2$-ice clouds and H$_2$O Earth-like clouds. This value is set to 0.5 when not otherwise specified. This choice is driven by the fact that the cloud fraction values adopted in \cite{biasiotti22} are meaningful only for the specific geographical configuration of the present time Earth and cannot be transposed easily to the conditions of the Early Mars. Using the same fractional cloud coverage across all models allows us to compare the results of different ESTM runs more straightforwardly. In Sect.~\ref{sss:clouds} we explored also the effect of different cloud coverages in the [0, 1] range.

\subsection{EOS} \label{ss:eos}

The OLR and TOA albedo lookup tables used in ESTM have been provided by the EOS code \citep{simonetti22}. EOS is based on the RT code HELIOS \citep{malik17,malik19} and the opacity calculator HELIOS-K \citep{grimm15,grimm21}, but tailored for the study of terrestrial-type atmospheres. EOS inherits the feature of being GPU-accelerated. It operates in the two-stream approximation with an improved non-isotropic scattering treatment \citep{heng18} and can carry on calculations either line-by-line or using k-distribution tables.

For this work we adopted the spectral line lists for CO$_2$, H$_2$O and CH$_4$ available on HITRAN2020 \citep{gordon22}. Calculations have been performed using k-distribution tables that we checked against previous EOS results obtained with the line-by-line procedure. The absorption of each gas has been calculated on a log-spaced grid in pressure with a 1/3 dex step in the interval between $10^{-6}$ and $10^3$ bar, and a linearly spaced grid in temperature with a 50 K step between 50 and 450 K. The spectral interval considered is 0-35000 cm$^{-1}$. The CO$_2$ absorption lines have been modeled following \cite{perrin89} and cutting the wings at 500 cm$^{-1}$ from the line centre, while H$_2$O and CH$_4$ lines have been modeled using a Voigt distribution cut at 25 cm$^{-1}$ from the line centre. For H$_2$O, the residual absorption at distances larger than 25 cm$^{-1}$ (the so-called Lorentzian pedestal) has been removed, since it is already accounted for in the collisional continuum. We then included the CO$_2$-CO$_2$ Collision-Induced Absorption (CIA) in the 0-750 cm$^{-1}$ \citep{gruszka97,gruszka98} and 1000-1800 cm$^{-1}$ \citep{baranov99} ranges \citep[the so-called GBB continuum used in][]{wordsworth10} and the CO$_2$-CH$_4$ CIAs in the 0-2000 cm$^{-1}$ range \citep{wordsworth17}. Concerning H$_2$O, we used the self- and foreign-induced continua calculated using the MT\_CKD v3.4 \citep{clough89,mlawer12} model available as a standalone subroutine of the LBLRTM code \citep{clough05}. It should be noted that the foreign-induced H$_2$O continuum is calculated for a modern Earth-like composition and thus it might deviate from that produced by a CO$_2$-dominated atmosphere. However, previous versions of this model have been used in other studies of the Early Mars atmosphere \citep[e.g.][]{wordsworth13}, thus at least guaranteeing some level of intercomparison. The atmosphere has been modeled as composed by 10 log-spaced layers per order of magnitude in pressure. Since adopting a coarser or finer pressure grid for the atmosphere impacts the EOS results \citep[][Fig.~2]{simonetti22}, we chose the total number of layers in order to
be as homogeneous as possible across different cases. The pressure at the upper face of the uppermost layer ($P_{\text{TOA}}$) is always $10^{-5}$ bar. Rayleigh scattering cross-sections for CO$_2$ and CH$_4$ \citep{sneep05} and H$_2$O \citep{wagner08} are also included. RT calculations are all clear-sky, since the effect of clouds are already included in ESTM (see Section \ref{sss:clouds}).

The adopted vertical atmospheric pressure-temperature (PT) profile consists of a convective troposphere and an isothermal stratosphere. This vertical PT profile considers the condensation of H$_2$O (if any) in the lower troposphere, the condensation of CO$_2$ (if possible) in the upper troposphere and an isothermal stratosphere at 155 K. The non-ideal behaviour of the involved gases is taken into account (see Appendix A). The resultant lapse rates are used both for RT calculations and the altimetric adjustment described in Section \ref{sss:topography}.

For each of the tested atmospheric composition and $P_0$ in the [0.1, 5.62] bar range we constructed two lookup tables, which are then used in the ESTM climate model: one containing the OLR as a function of $T_0$, and another containing the TOA albedo as a function of $T_0$, the surface albedo $\alpha_s$ and the stellar zenith angle $z$. These have been obtained by running EOS for each combination of the relevant parameters. The temperature grid is linearly spaced, with a resolution of 10 K in the 160-240 K range and a resolution of 20 K in the 240-360 K range, the zenith angle grid is $\{0^\circ, 45^\circ,\, 60^\circ,\, 70^\circ,\, 80^\circ,\, 85^\circ,\, 88^\circ\}$ and the surface albedo grid is $\{0,\, 0.30,\, 0.45,\, 0.60,\, 0.90\}$. Some of the $(T_0,P_0)$ points are associated with unphysical conditions, i.e. positions in the pressure-temperature plane that resides below the condensation curve of CO$_2$. In that case, we substituted the surface pressure value taken from the pressure grid with the CO$_2$ condensation pressure at that temperature. As an example, the OLR and TOA albedo in the $(200\,\text{K},2.0\,\text{bar})$ point has actually been calculated using 1.578 bar. This choice allows us to self consistently take into consideration the radiative impact of atmospheric condensation at surface. The OLR for $T_0 \le 150$ K is calculated from the Stefan-Boltzmann law, and is linearly interpolated between 150 K and 160 K. This is reasonable considering that the CO$_2$ condensation pressure at 150 K is 0.008 bar and thus the expected atmospheric radiative forcing is small. 
For $T_0 < 160$ K the TOA albedo depends on the zenith angle and the surface albedo, but not on $T_0$, since below that temperature a CO$_2$-dominated atmosphere must be thin due to condensation.



\subsubsection{Atmospheric chemical composition} \label{sss:atmospheres}

The atmospheres tested in this work are all CO$_2$-dominated. This is a standard choice for the study of Early Mars \citep[e.g.][]{kasting91} and it is justified by considerations on the chemical composition of the outgassed fluids \citep{grott11}. At present time, the martian atmosphere includes a 5\% by volume of N$_2$ and some modelers \citep[e.g.][]{ramirez14} scaled this abundance in their models. Anyway, our choice to not include N$_2$ has only negligible effects on the radiative properties of the atmosphere, as it is shown in \cite{ramirez17b}. Fig.~\ref{fig:altimetry_and_avgtemp}, panel (b), confirms this result: substituting 5\% of CO$_2$ with N$_2$ in our model would reduce $\langle\overline{T}_0\rangle$ by $\sim$2 K at maximum, and by a much smaller quantity in most other cases.

We also included a variable amount of H$_2$O, by fixing the relative humidity $RH$ to either 0\%, 20\% or 40\%. Since our RT lookup tables must be representative of the average atmospheric mixture, we chose low $RH$ values that are representative of the typical humidities found in modern Earth deserts \citep{yang20}. More specifically, we associated the null $RH$ value to the model with no oceans, the intermediate $RH$ value to the models with the 150-m GEL ocean and the highest $RH$ value to the model with the 550-m GEL ocean.

We then tested three different fractional amounts of CH$_4$, $f_{\text{CH4}}$: 0, 0.1\% and 1\% by volume. There are two reasons behind these specific choices. First, low amounts of CH$_4$ prevent the formation of organic hazes via photochemical reactions in the upper atmosphere. These hazes form efficiently when the CH$_4$/CO$_2$ ratio increases above $\sim 0.1$ \citep{trainer06,dewitt09}. Second, low amounts of CH$_4$ are possibly easier to explain in light of the known geochemical features of Mars. In general, the amount of CH$_4$ in the Early Mars atmosphere is understood to be primarily controlled by the balance between serpentinization\footnote{The process in which mafic and ultramafic rocks are oxidized at relatively low temperatures, i.e. in the planetary crust. This process sequester water and releases reduced species, such as CH$_4$.} of the olivine crust \citep{chassefiere13} and photo-dissociation \citep{wordsworth17} in the atmosphere. The photodissociation rate is limited by the availability of photons with $\lambda < 160$ nm, most of which are Lyman-$\alpha$ photons, whose flux at the Noachian-Hesperian boundary was about $3 \times 10^{15}$ m$^{-2}$s$^{-1}$ \citep{wordsworth17}. In few bars-level CO$_2$ dominated atmospheres, this limit is reached for $f_{\text{CH4}} \sim 1\%$ \citep{zahnle86}. If we approximate the photodissociation rate as linearly dependent on concentration when it is below this limit, we can make a zeroth-order estimate of the flux needed to maintain a given atmospheric concentration. On the basis of the photon flux, the source flux required to support a 1\% CH$_4$ concentration would be $3 \times 10^{15}$ m$^{-2}$s$^{-1}$, or one order of magnitude smaller for a 0.1\% concentration.
This quantity can be compared with the flux from regions undergoing serpentinization, that has been estimated to be in the $10^{16}-10^{18}$ m$^{-2}$s$^{-1}$ interval \citep{etiope13}. Taking the lower end of this interval, the methanogenic regions must cover 30\% of the martian surface to maintain our highest tested concentration. While this estimate is extremely simplified, it allows to check the plausibility of our choices.

\section{Results} \label{sec:results}

\begin{table}[]
\centering
\begin{tabular}{lcccc}
\hline
Name & RH & $f_{\text{CH4}}$ & Cloud & GEL \\
\quad & [\%] & [\%] & model & [m] \\
\hline
A0F0 & 0 & 0 & CO$_2$-ice & 0 \\
B0G0 & 20 & 0 & Earth-like & 0 \\
B0GH & 20 & 0 & Earth-like & 150 \\
BLG0 & 20 & 0.1 & Earth-like & 0 \\
BLGH & 20 & 0.1 & Earth-like & 150 \\
BHG0 & 20 & 1 & Earth-like & 0 \\
BHGH & 20 & 1 & Earth-like & 150 \\
C0G0 & 40 & 0 & Earth-like & 0 \\
C0GN & 40 & 0 & Earth-like & 550 \\
\hline
\end{tabular}
\caption{A summary of the models studied in this paper. The first letter refers to the relative humidity ("A" for 0\%, "B" for 20\%, "C" for 40\%). The second letter refers to the fractional amount of CH$_4$ ("0" for 0\%, "L" -low- for 0.1\%, "H" -high- for 1\%). The third letter refers to the cloud model ("F" for CO$_2$-ice, "G" for Earth-like). The fourth and last letter refers to the surface water distribution and depth ("0" for no surface water, "H" for an Hesperian-like 150-m GEL, "N" for a Noachian-like 550-m GEL).}
\label{tab:models}
\end{table}

Our goal is to explore the early martian climate over a wide range of planetary and atmospheric parameters. In particular, we are interested in studying which set of inputs enables localized, seasonal thaws under Noachian and Hesperian Mars conditions.

The models tested in this paper are summarized in Table \ref{tab:models}. For each of them we varied (i) the datum pressure $P_0$ between 0.1 and 5.62 bar in logarithmically spaced steps of 1/8 of dex and (ii) another parameter chosen among the obliquity $\varepsilon$, the eccentricity $e$, the land albedo $\alpha_l$ and the cloud fraction $f_c$. We also explored the interplay between obliquity, eccentricity and the argument of the perihelion $\omega$, for a fixed surface pressure value. We refer to a specific combination of input variables for a given model as a ``case'' and the total number of cases run in this study is $\sim$10$^4$.

In practice, we followed the steps described below. First, we run ESTM until convergence is found for a given set of input parameters. The resulting seasonal-latitudinal temperature map, $T_0$, refers to the altimetric datum and not to the actual surface. Second, we correct the seasonal-latitudinal $T_0$ map of ESTM for the altitude as described in Section \ref{sss:topography}, obtaining $T_s$. Third, we search for the highest ($T_{\text{max}}$) and lowest ($T_{\text{min}}$) temperature in the $T_s$ map. If $T_{\text{max}}$ is above the freezing temperature of water, then we calculate the fraction of the martian year for which this condition 
is satisfied at that specific latitude. $T_{\text{max}}$ imposes the most stringent limit on the melting of ices at a given latitude. Fourth, we check which latitude lasts the most above-freezing temperatures and save the result as a fraction of martian year. Usually, but not always, it is the same latitude at which $T_{\text{max}}$ occurs. In particular, we considered two different threshold temperatures $T_{\text{thr}}$ as "above-freezing": 273 K and 252 K. The first one refers to pure water, the second one refers to a saturated solution of NaCl \citep[i.e. brine,][]{fairen10}. These thresholds are reported in Figs.~\ref{fig:results_obliquity}-\ref{fig:results_orbits} as red and yellow lines, respectively. Finally, using $T_{\text{min}}$, we check the run for CO$_2$ condensation at surface, which signals the potential beginning of an atmospheric collapse. Notice that we do not compute the duration or magnitude of this condition: even if condensation occurs at a single latitude point and for a single time-step interval ($\sim$14 d), that run will still be flagged as "condensing". In Figs.~\ref{fig:results_obliquity}-\ref{fig:results_orbits}, the regions of the parameter space in which we found condensation of the atmospheric CO$_2$ at surface are hatched. Notice that $T_{\text{max}}$ and $T_{\text{min}}$ generally occurs at different latitudes and positions along the orbit.

All runs start using an isothermal latitudinal profile at 300 K. Thus, no ocean ice is present at the initial time of the simulation. Due to climate bistability, the final climate state 
might depend on the starting temperature. However, since: (i) the only source of bistability in ESTM is the ice-albedo feedback \citep{murante20}, (ii) only ocean ices are tracked dynamically and (iii) the ocean coverage is small, we expect bistability to be generally negligible. In Appendix \ref{app:profiles}, we show and discuss the latitudinal OLR, TOA albedo and diffusion profiles for a small number of selected cases.

\subsection{Dependence on obliquity}\label{ss:obliquity_results}

\begin{figure*}
\plottwo{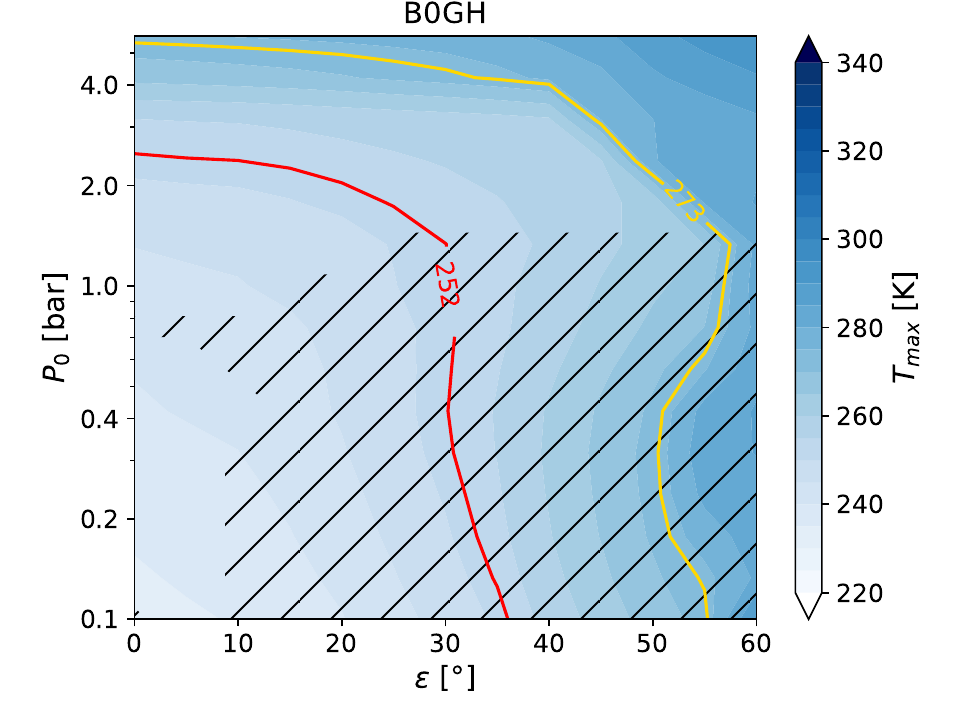}{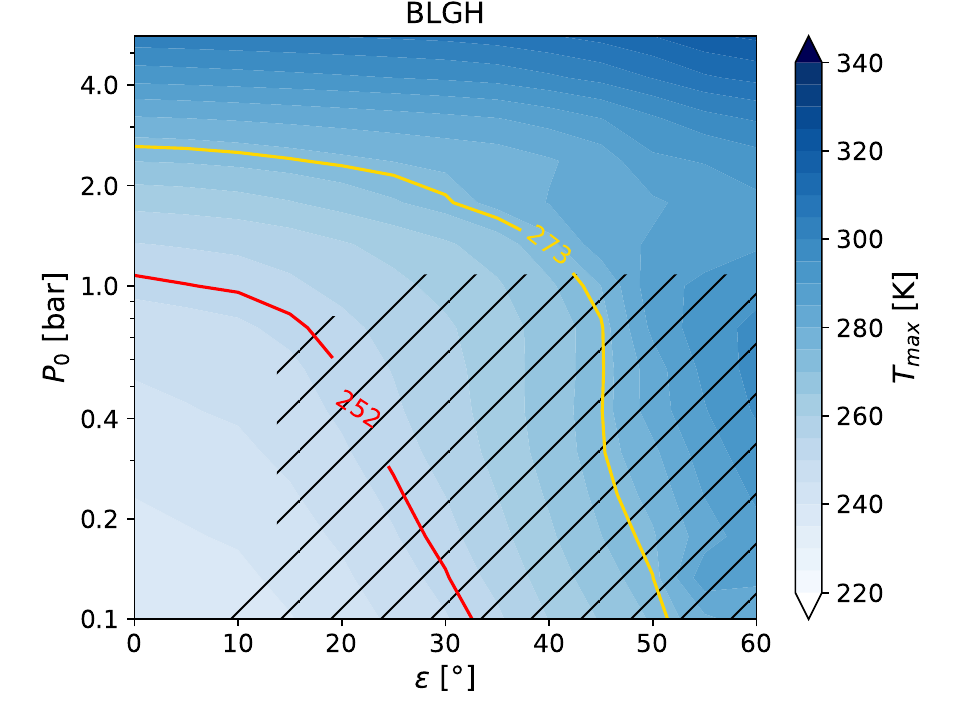}
\plottwo{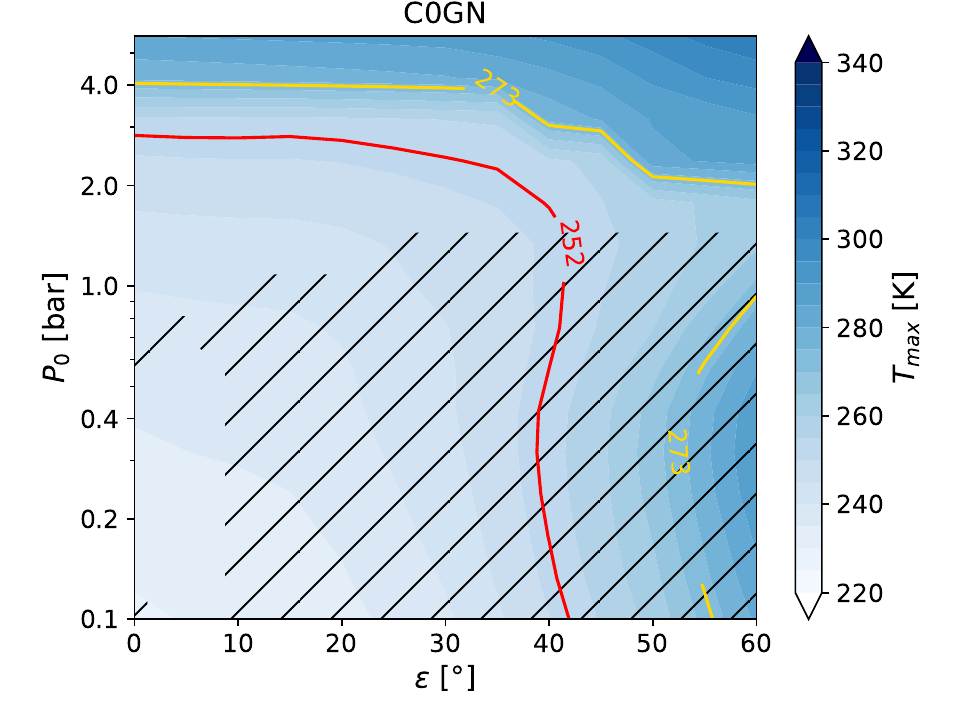}{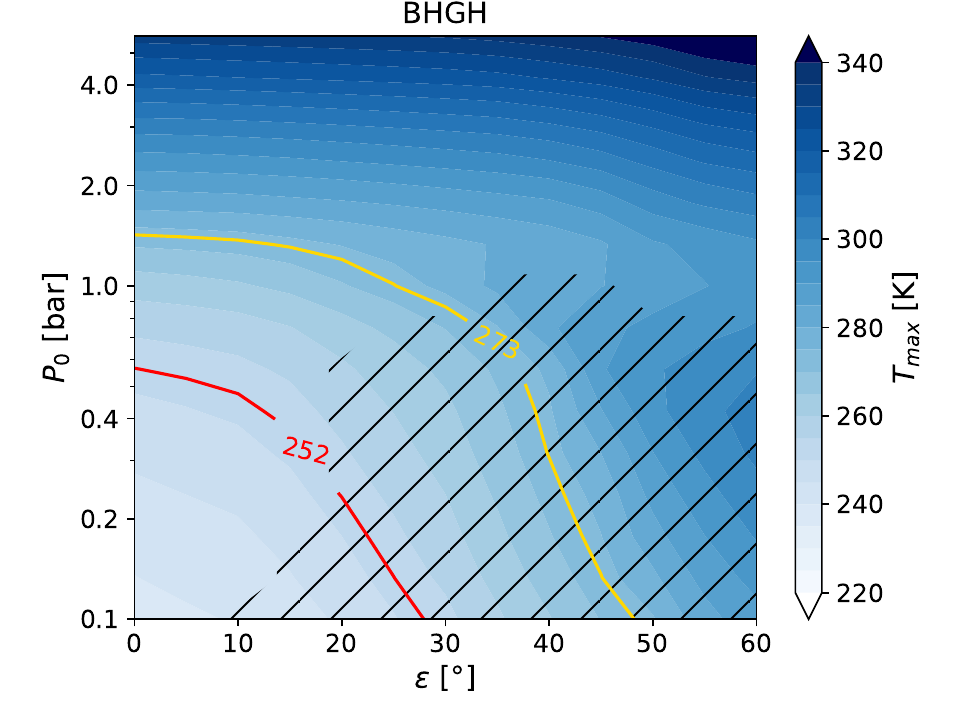}
\caption{The maximum surface temperature $T_{\text{max}}$ as a function of the surface pressure and obliquity. From top left moving clockwise: maps for the B0GH, BLGH, BHGH, C0GN models. Yellow and red contour lines highlight the regions of the parameter space for which pure liquid water and brine can be liquid, respectively. Hatched regions refer to combinations of parameters that causes the CO$_2$ in the atmosphere to condense at surface.
\label{fig:results_obliquity}}
\end{figure*}



We explored the range of obliquity values between 0$^\circ$ and 60$^\circ$, sampled at 5$^\circ$ intervals. These boundaries have been chosen considering the results of \cite{laskar04}, whose calculations showed that the probability of an obliquity excursion up to 60$^\circ$ in a span of 250 Myr is around 45\%. The results for the models B0GH, BLGH, BHGH and C0GN are shown in Fig.~\ref{fig:results_obliquity}.


All models exhibit a similar structure, with a low pressure ($< 1.0$ bar), mid-to-high obliquity ($> 10^\circ$) region of the parameter space in which atmospheric condensation at surface occurs at some moment along the orbit. The limiting pressure that identifies this region is particularly easy to identify, since no case with 1.78 bar or more in models B0GH and C0GN and no case with 1.33 bar or more in models BLGH and BHGH is predicted to have surface condensation, independently of the obliquity. 

In general,  $T_{\text{max}}$ is roughly independent of pressure when $P_0 < 1.0$ bar, and roughly independent of obliquity when $P_0 > 4.0$ bar. Between these two values there is a transition region in which both parameters influence the outcome. The low pressure region shows some non-linear behaviour, since all models exhibit a local maximum of $T_{\text{max}}$ for $P \sim 0.4$ bar and $\varepsilon = 60^\circ$.

Both the transition from obliquity-dominated to pressure-dominated $T_{\text{max}}$ and the low-pressure local maximum are caused by the interplay between the increasing heat redistribution efficiency and the increasing greenhouse effect of a thicker atmosphere. Between 0.1 and 0.4 bar, $T_{\text{max}}$ likely increases because the stronger infrared absorption overcomes the more efficient redistribution of heat, that tends to smooth out the seasonal and latitudinal differences. Above 0.4 bar, instead, the diffusion term becomes more important, leading to lower $T_{\text{max}}$. Finally, above 1.33 bar, $T_{\text{max}}$ starts increasing again because the entire planet becomes warmer, while latitudinal variations are suppressed even at high obliquities.

In model C0GN, there is a rapid transition in the $T_{\text{max}}$ values near the isotherm at 273 K. This transition is present also in the other models but it is less evident, and nearly disappears in the BHGH case. 
This behaviour is caused by the transition from a completely and continuously glaciated northern ocean, and a mostly open ocean. Oceans have a very low albedo even compared to martian regolith \citep[$\sim$0.05\footnote{For $z\sim60^\circ$.} vs 0.215, see e.g.][for data on the former]{huang19}, thus they contribute to warm up the planet. Since the ocean fraction is highest in the C0GN model, their effect is strongest.

We do not show the results for the A0G0 model because nearly all the tested obliquity-pressure combinations caused the atmosphere to collapse at the surface. More specifically, only the cases with $P_0 =$ 0.13 or 0.18 bar and $\varepsilon=5^\circ$ are stable, and their $T_{\text{max}}$ is equal to 236.9 and 237.9 K respectively. In the A0F0 model the maximum greenhouse limit is reached at around 2 bar (see Fig.~\ref{fig:altimetry_and_avgtemp}, panel b, red line), which causes the disappearance of the conditions conductive to seasonal thaws at higher pressures.


\subsection{Dependence on eccentricity}\label{ss:eccentricity_results}

\begin{figure*}
\plottwo{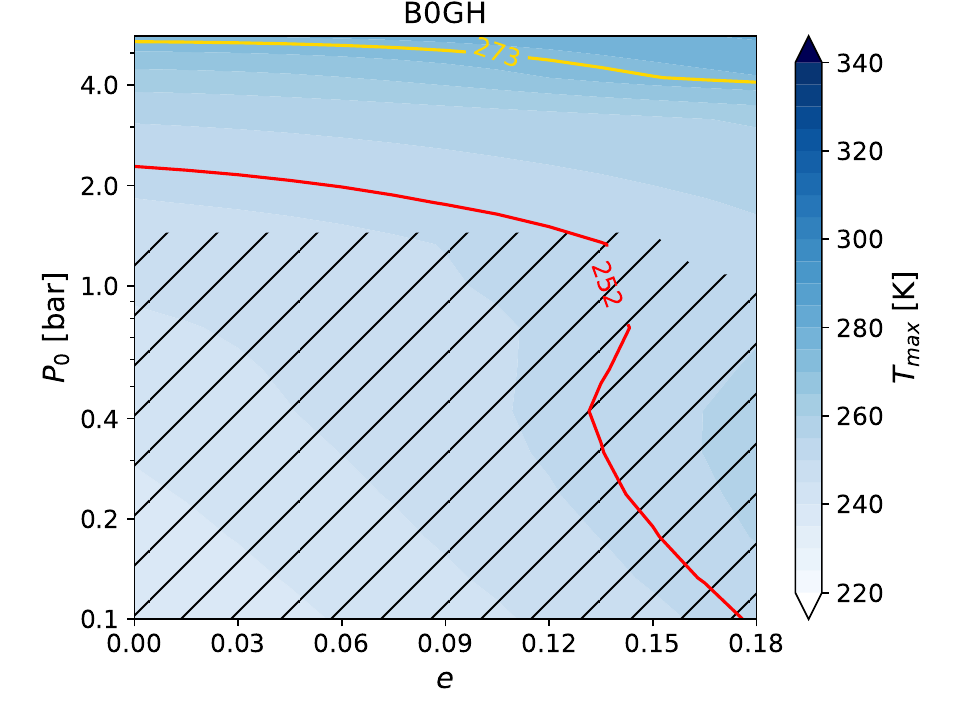}{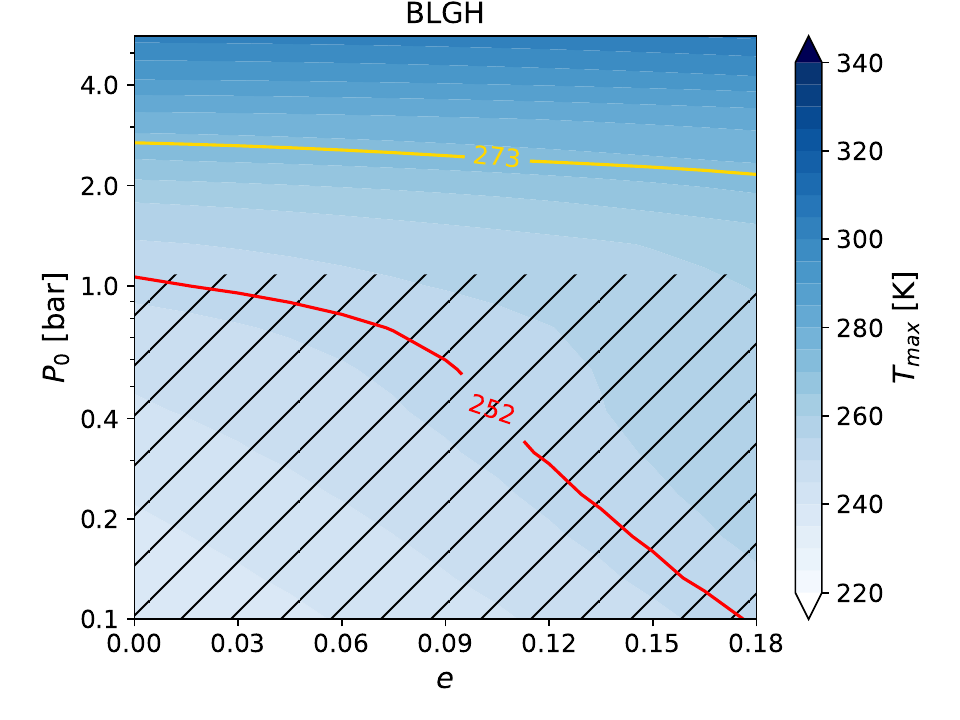}
\plottwo{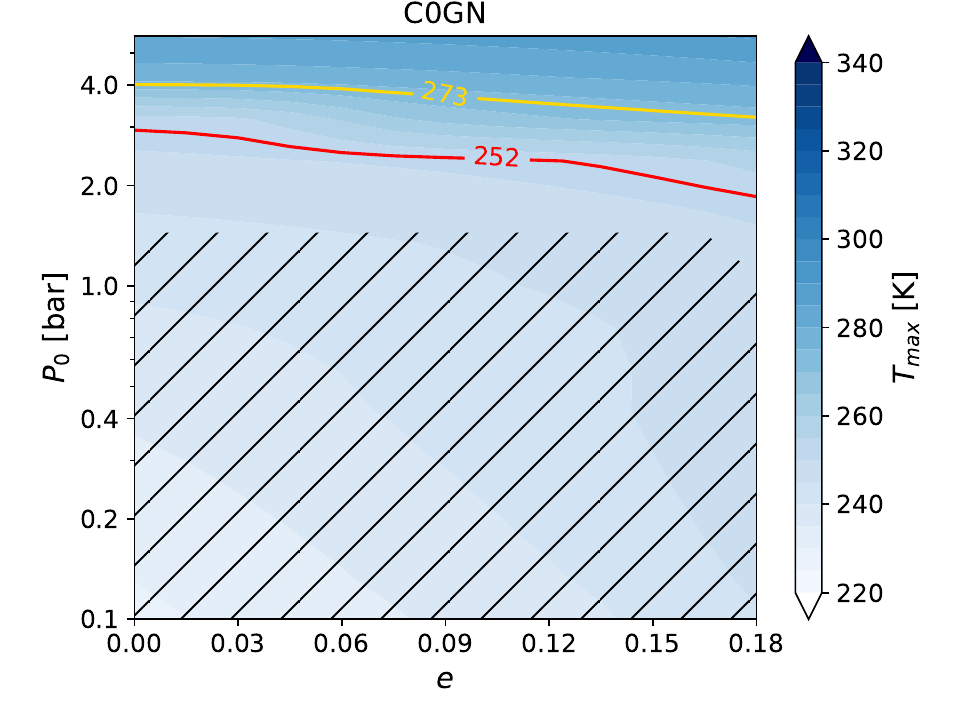}{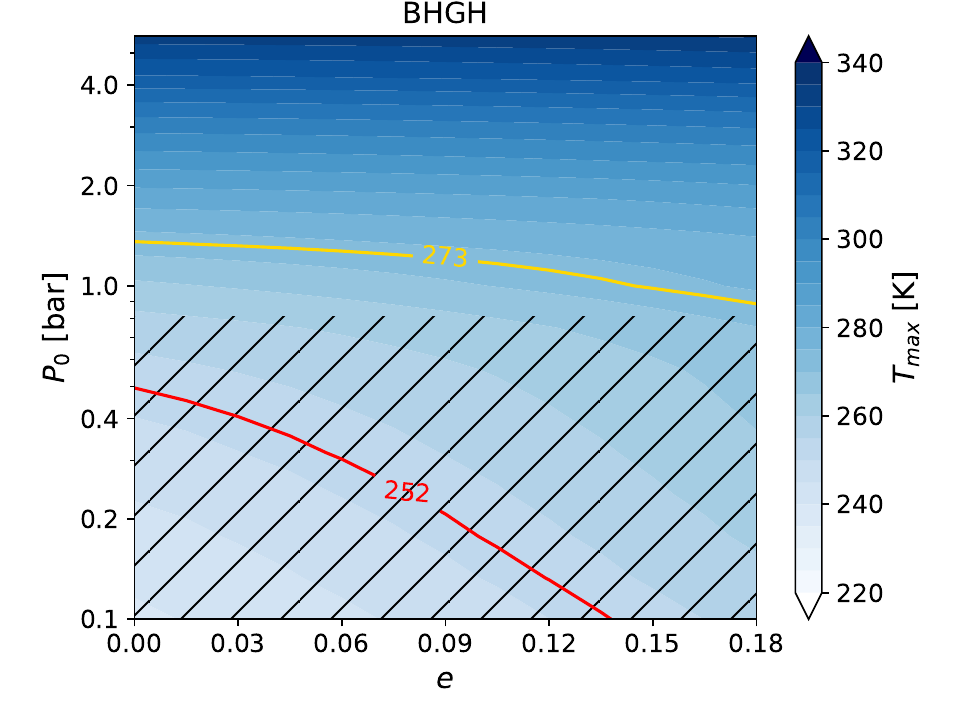}
\caption{The maximum surface temperature $T_{\text{max}}$ as a function of the surface pressure and eccentricity. From top left moving clockwise: maps for the B0GH, BLGH, BHGH, C0GN models. \label{fig:results_eccentricity}}
\end{figure*}

The second parameter we investigated is the orbital eccentricity, for which we explored the range of values between 0 and 0.18, in 0.015 steps. Again, the boundaries were informed by the work of \cite{laskar04}, which predicted a potential eccentricity excursion up to $\sim 0.2$ in their 5 Gyr runs. At variance with the obliquity interval studied in the previous Section, eccentricity values above 0.14 have a low ($<5$\%) occurrence probability in a 250 Myr time span and are thus unrealistic, even though we included them for completeness.

Eccentricity variations have two effects on the surface temperature of the planet. First, increasing eccentricity increases the average instellation of the planet by a factor $(1-e^2)^{-1/2}$ \citep{williams97}. Second, the higher eccentricities contribute to seasonal variations, with the effect being strongest when $\omega=90^\circ$ or $270^\circ$ and weakest when $\omega=0^\circ$ or $180^\circ$\footnote{If $\omega=90^\circ$, the planet is at periastron during the southern hemisphere summer solstice, meaning that the seasonal variations due to obliquity and eccentricity compound together. The same happens if $\omega=270^\circ$, but considering the northern hemisphere summer solstice. Instead, when $\omega=0^\circ$ or $180^\circ$, the planet is at equinox when it passes through the periastron.}. In the models we are considering here, the first effect is small since going from 0 to 0.18 increases the martian insolation by 1.7\%, while the second is not sufficient to influence the final results, despite that $\omega=286.5^\circ$. This is evident observing the results in Fig.~\ref{fig:results_eccentricity}: most of the variation happens on the vertical axis, i.e. when we changed the surface pressure. At low pressures ($<1$ bar) there is a modest increase of $T_{\text{max}}$ for higher eccentricities, but the limiting pressure required for thawing, under the more conservative condition, changes only by $\sim 30\%$.

Seasonal variability due to obliquity is dominant also because it sets the limiting $P_0$ to avoid atmospheric collapse at the surface. $P_0$ is equal to 1.78 bar for models B0GH and C0GN, 1.33 bar for model BLGH and 1 bar for BHGH. In models B0GH and C0GN this limit is reduced to 1.33 bar for the highest eccentricity cases, namely those above 0.15.

Once again, we do not show the results for the A0F0 models, since for all the tested combinations of pressure and eccentricity the atmosphere is unstable against surface condensation. This is expected, since seasonality is dominated by obliquity and for $\varepsilon=25^\circ$ there are no non-condensing cases (see the previous Section). The highest $T_{\text{max}}$, just a few kelvins above the freezing temperature of pure water, are found for $e>0.15$ and $0.13<P_0<0.421$, and as for the other models, most of the variations are caused by pressure changes.

\subsection{Dependence on land albedo}\label{ss:surfalbedo_results}

\begin{figure*}
\plottwo{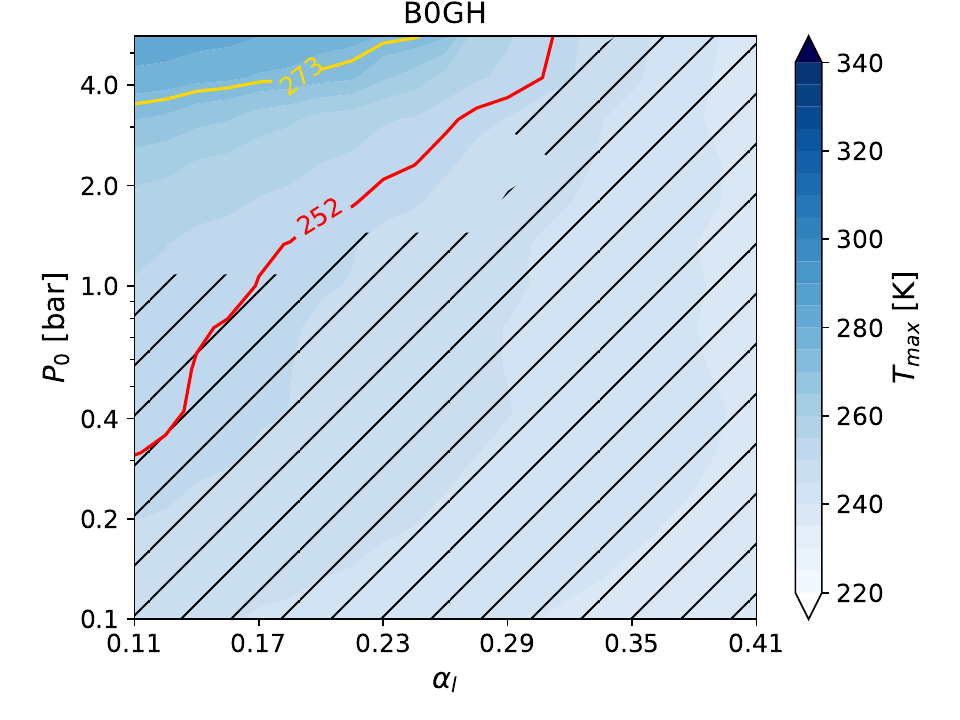}{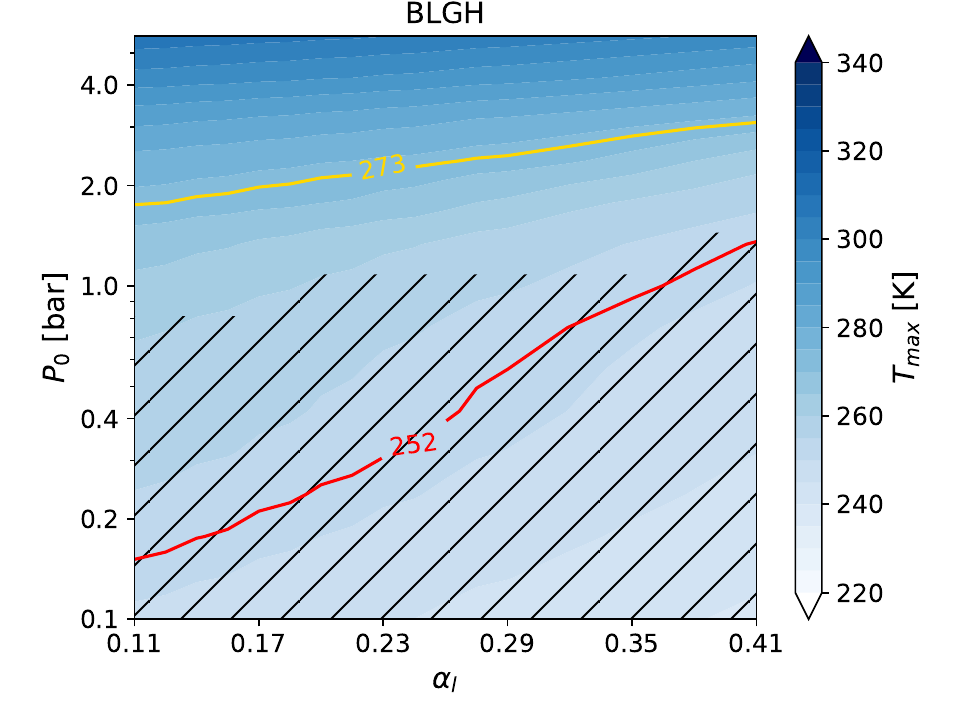}
\plottwo{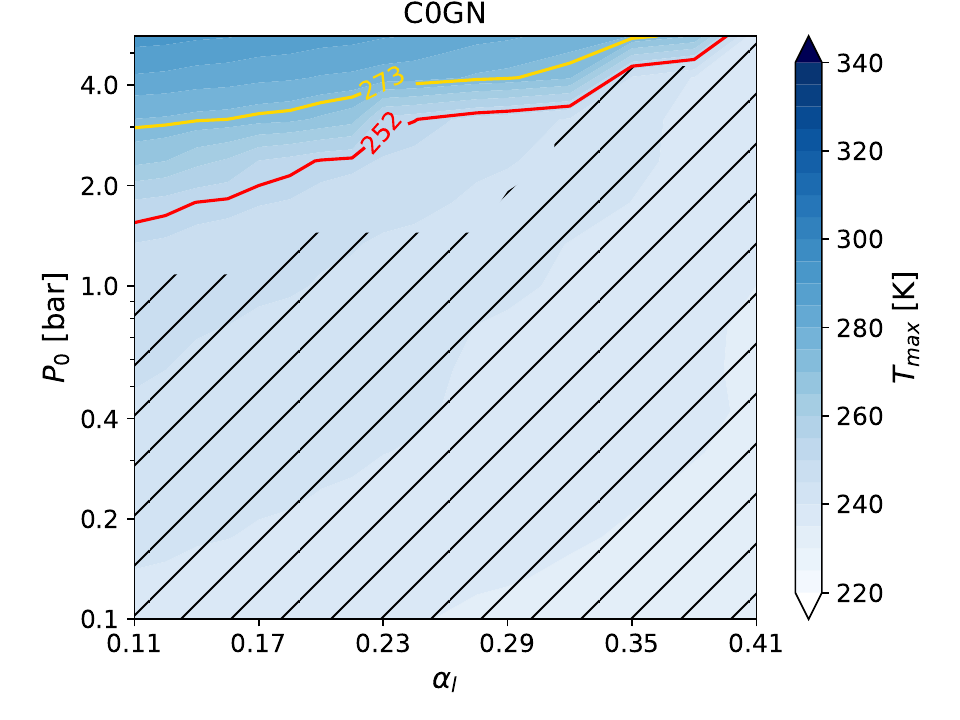}{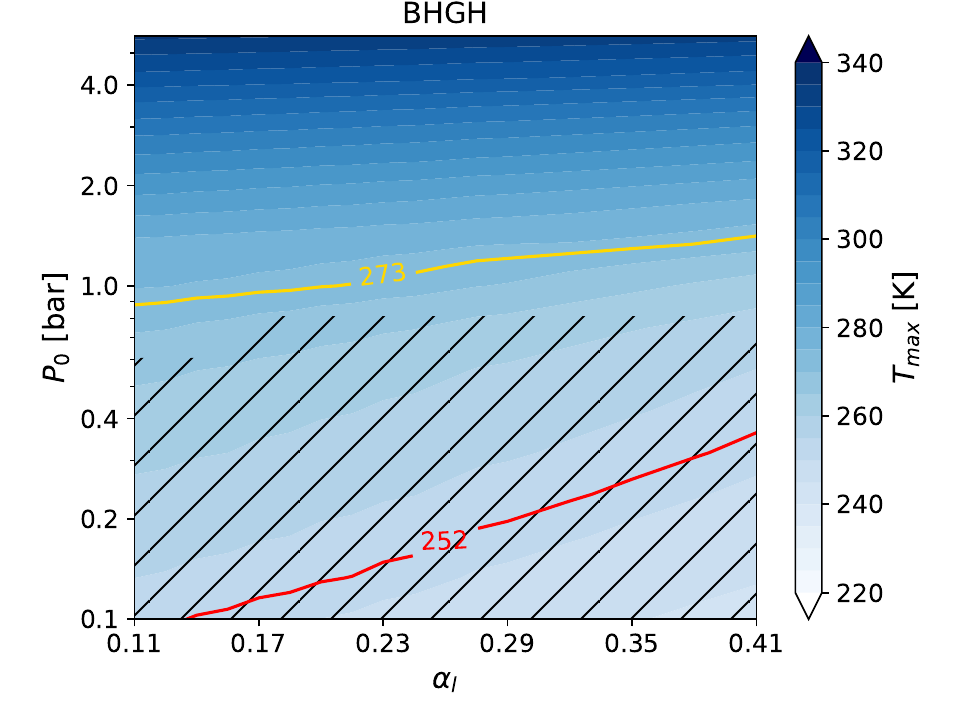}
\caption{The maximum surface temperature $T_{\text{max}}$ as a function of the surface pressure and the land albedo. From top left moving clockwise: maps for the B0GH, BLGH, BHGH, C0GN models. \label{fig:results_surfalbedo}}
\end{figure*}

The third parameter tested is the land albedo. Currently, its average value is 0.215, but during the Noachian and Hesperian its value could have been different. Two factors that can change the land albedo are: (i) large volcanic eruptions that cover a substantial fraction of the surface with fresh basalt, whose albedo is lower than that of regolith, and (ii) reflective permanent glaciers in the southern highlands. In this work, we varied the land albedo between 0.11 and 0.41 in 0.015 steps. The lower end is meant to represent the albedo of surface mostly covered in basalt quenched glass \citep[whose albedo is 0.09,][]{essack20}, while the upper end is meant to represent a surface which is 50\% covered in modern Mars regolith and 50\% covered in ice (albedo: 0.6), or an Earth-like dry sand.

The results of this exploration are shown in Fig.~\ref{fig:results_surfalbedo}. The effect of changing the land albedo is stronger in models B0GH and C0GH and weaker in models BLGH and BHGH, and decreases when $P_0$ increases. In models B0GH and C0GN, high land albedos prevent seasonal deglaciations both using the more conservative and the more lenient condition, even at the highest pressure tested. Samely, they favor the collapse of the atmosphere when higher than 0.32 (for model B0GH) and 0.38 (for model C0GN). On the other hand, low albedo runs still require at least 1 bar of CO$_2$ both in B0GH and C0GN in order to prevent atmospheric collapse. As in the other set of calculations, the C0GN model shows a faster transition in $T_{\text{max}}$ around the 273 K isoline, which again underlines the transition from a completely frozen northern ocean to a mostly open one, and it is more evident than in the other models because this ocean covers a larger fraction of the surface. This transition is steeper for high land albedos due to the higher albedo contrast between light lands and dark oceans.

The models BLGH and BHGH allows instead a stable atmosphere and seasonal thaws in the northern lowlands even for the highest land albedo value tested, provided the surface pressure is sufficiently high. In model BLGH the limiting $P_0$ is 3.16 bar and in model BHGH it is 1.33 bar. On the other hand, the A0F0 model (not shown) can neither avoid surface atmospheric condensation nor produce seasonal thaws (under the pure water condition), even at the lowest land albedo tested.

\subsection{Dependence on the cloud fraction}\label{ss:cloud_results}

\begin{figure*}
\plottwo{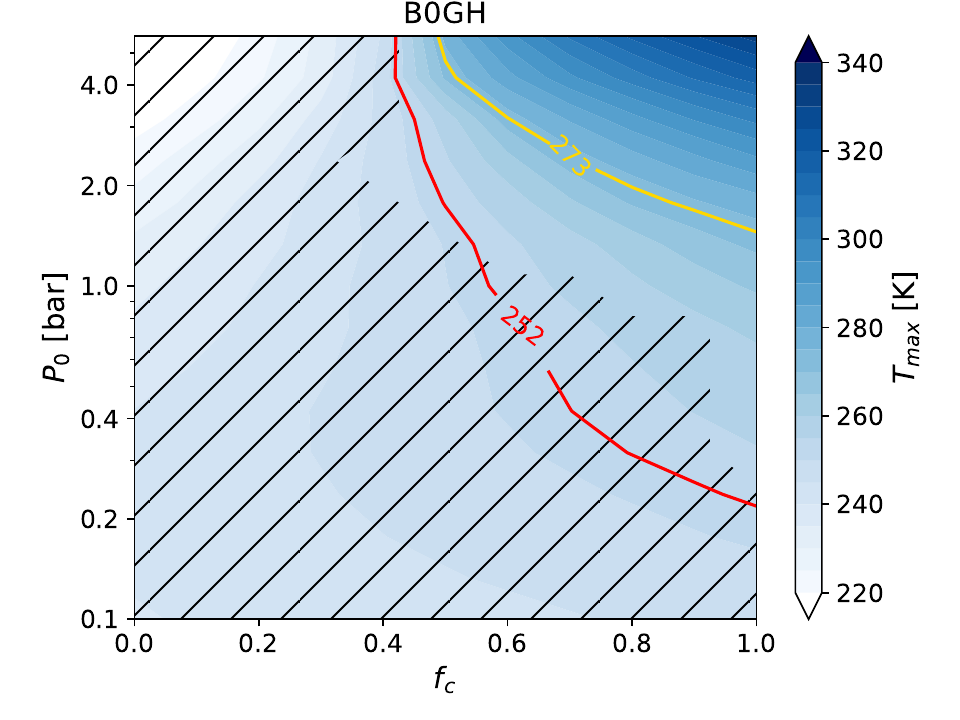}{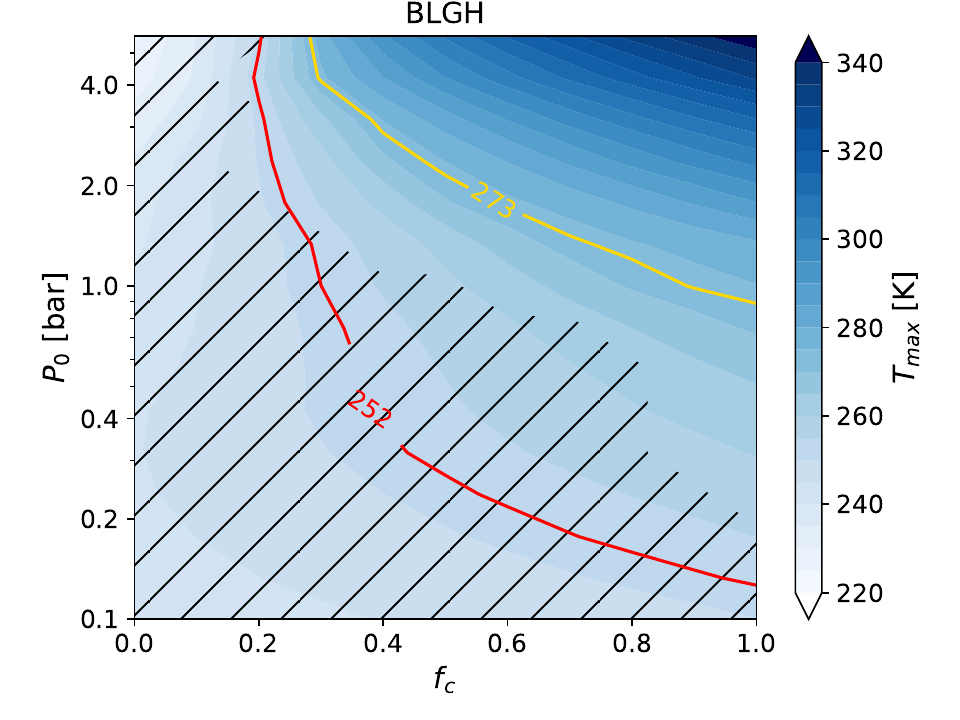}
\plottwo{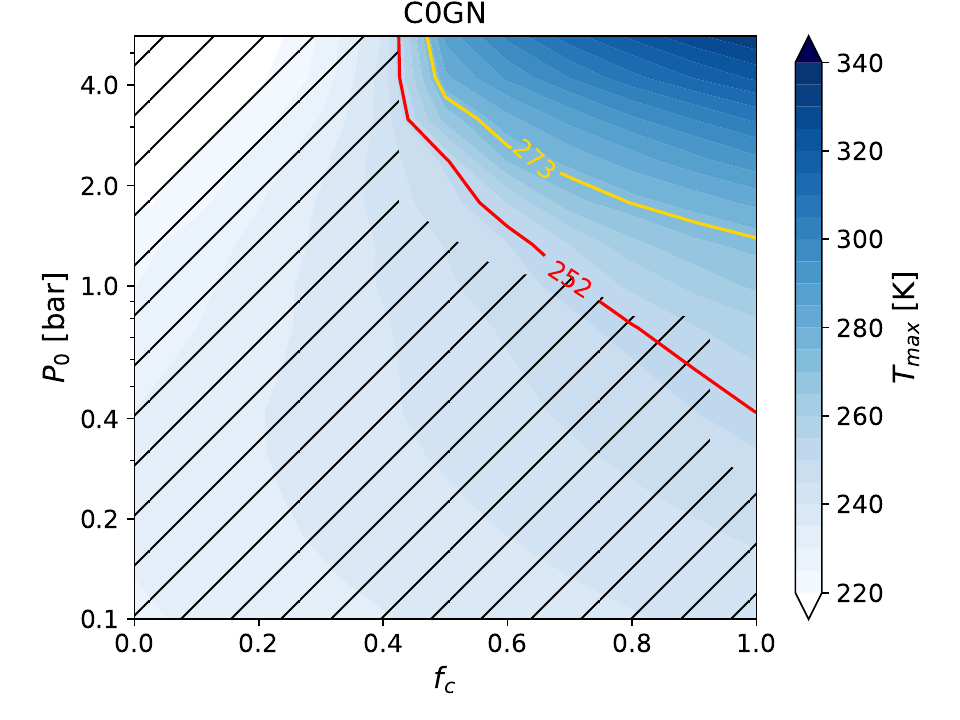}{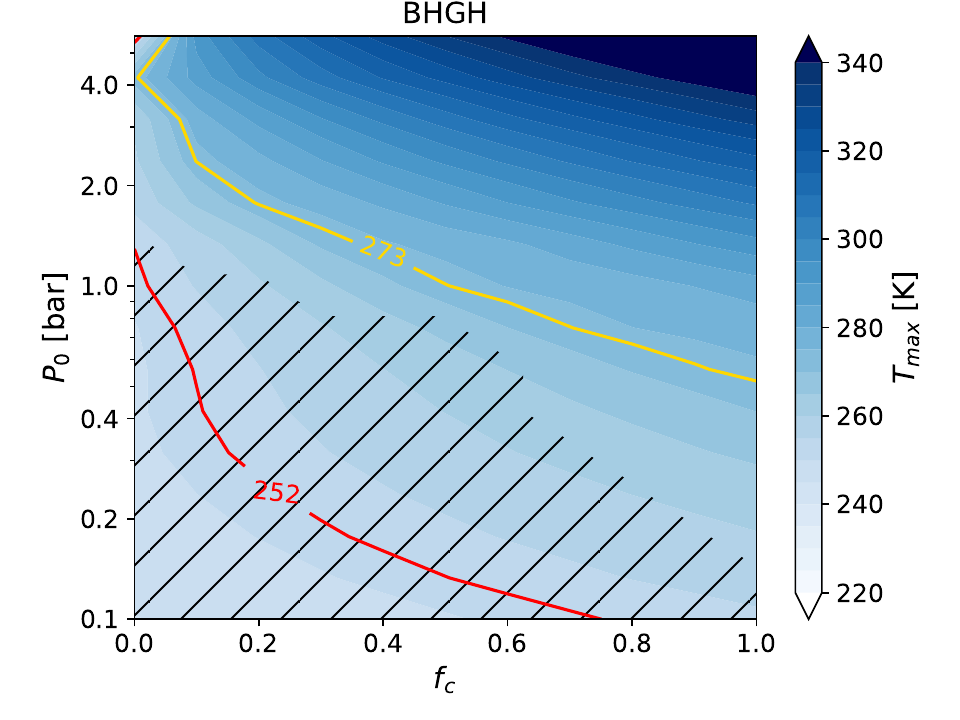}
\includegraphics[width=0.42\textwidth]{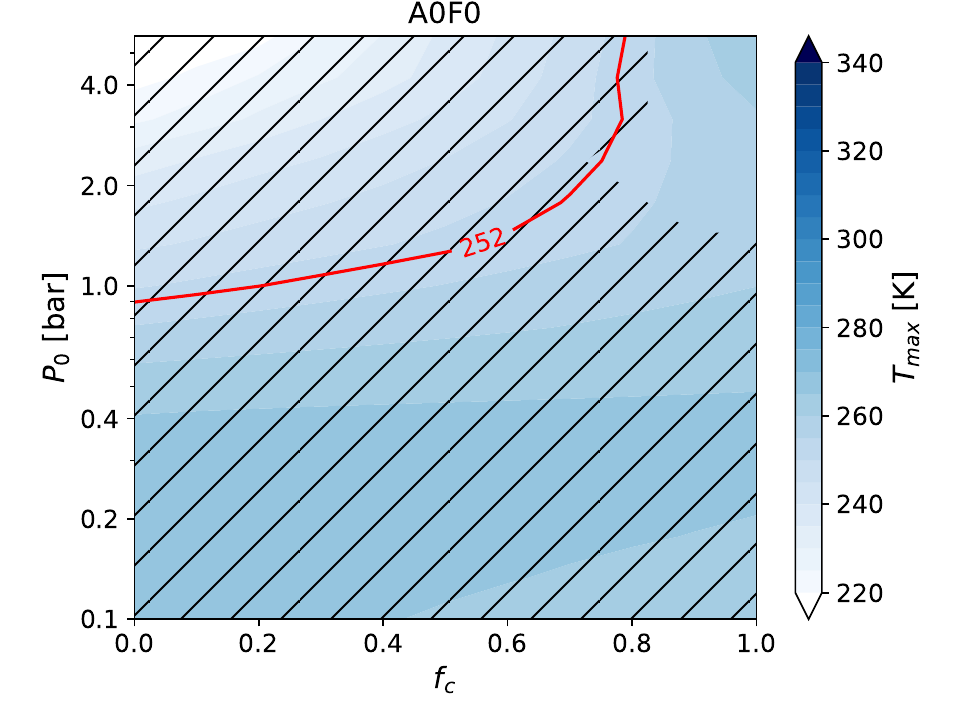}
\caption{The maximum surface temperature $T_{\text{max}}$ as a function of the surface pressure and the cloud fraction. From top left moving clockwise: maps for the B0GH, BLGH, BHGH, A0F0, C0GN models. We recall that in models B0GH, BLGH, BHGH and C0GN we use Earth-like H$_2$O clouds, while in model A0F0 we use CO$_2$ ice clouds. \label{fig:results_clouds}}
\end{figure*}

The fourth and final parameter considered is the cloud cover fraction. Modeling the cloud coverage and its effects in conditions different from that of the modern Earth is a notoriously difficult task, and different GCMs can produce substantially different results for the same input \citep[see e.g.][]{sergeev22}. In absence of a self-consistent model for clouds, we wanted to explore the implications of the simple parameterization presented in Sect.~\ref{sss:clouds} by varying the cloud cover fraction between 0\% and 100\%, in 10\% steps. Alternatively, this analysis can be intended as studying how, for a fixed but unknown coverage, the climate responds when our simplified cloud properties are linearly scaled.

The results are shown in Fig.~\ref{fig:results_clouds}. Since we tested two different cloud parameter sets (see Tab.~\ref{tab:clouds}), it is more meaningful to treat the models B0GH, BLGH, BHGH and C0GN separately from the model A0F0. 

\subsubsection{H2O clouds}\label{sss:cloud_earthlike_results}

Concerning the cases for which we adopted Earth-like H$_2$O clouds, any increase in cloud fraction causes $T_{\text{max}}$ to increase. In other words, this type of clouds are net greenhouse contributors across all the tested pressure range. A direct consequence of this fact is that, in general, we need at least some cloud coverage to avoid the atmospheric collapse and the presence of seasonal thaws. The minimum coverage value for a stable atmosphere is between 40\% and 50\% for B0GH and C0GN and between 10\% and 20\% for BLGH. The BHGH model always allows for both non-condensing and thawing conditions, irrespective of the cloud coverage. All models see their atmospheric stability region widened when the cloud fraction value is higher. At 100\% coverage, all the models with Earth-like H$_2$O clouds are stable down to 0.32 bar (and BHGH down to 0.13 bar).

As can be seen from  Fig.~\ref{fig:results_clouds}, the net greenhouse effect produced by clouds is also strongly dependent on $P_0$, which in turn causes some interesting secondary effects. First of all, at variance with all the previously tested parameters, high pressure ($>4$ bar) cases have not always a hotter summer than lower pressure ones. Apart for the BHGH model, the minimum $T_{\text{max}}$ value is found at the maximum tested pressure, for a 0\% coverage. In the BHGH model, the minimum $T_{\text{max}}$ is encountered at 0.1 bar (244 K) as in the previous explorations. However, $T_{\text{max}}$ does not monotonically increase with pressure, and after reaching a maximum value equal to 272 K it decreases again to 247 K at $P_0=5.62$ bar. Our results indicate that at high pressures, the net effects of clouds is the strongest, contributing up to $\sim$150 K to $T_{\text{max}}$ (in BHGH), while for the lower pressures, the effect is as small as 7 K (in B0GH). A second interesting effect is that, in all models other than BHGH, there exist a cloud fraction value that makes $T_{\text{max}}$ insensitive to $P_0$. This value is around 40\% for B0GH and C0GN and around 20\% for BLGH. This means that variability on the vertical axis of all the models shown in this work depends at least partially on the exact choice of the cloud parameters, and that the stronger variability shown by the BLGH and BHGH models in Figs.~\ref{fig:results_obliquity}, \ref{fig:results_eccentricity} and \ref{fig:results_surfalbedo} is at least partially linked to the specific cloud coverage that we chose. Decreasing the cloud coverage, i.e. approaching the clear-sky case, reduces the variability as a function of $P_0$ for all models, while the opposite is true when we increase it.

\subsubsection{CO2 clouds}\label{sss:cloud_forget_results}

The behavior of $T_{\text{max}}$ as a function of cloud coverage in the model with CO$_2$ ice clouds is somewhat different. First, this type of clouds have a minor cooling effect at low pressure. As can be seen in the lower center panel of Fig.~\ref{fig:results_clouds} (model A0F0), going from 0\% to 100\% coverage, $T_{\text{max}}$ decreases from 267 K to 262 K at 0.1 bar and are substantially neutral at 0.32 bar. At 5.62 bar, they instead contribute $\sim 50$ K at 100\% coverage. The effect of clouds on $T_{\text{max}}$ is thus different from the one on the average planetary temperature reported in Fig.~\ref{fig:oceans_and_clouds}, panel (b). For pressures above 1.33 bar, the greenhouse contribution of clouds is sufficient to prevent the atmospheric collapse, provided the coverage is above 70\%. However, even at 100\%, $T_{\text{max}}$ is not sufficient for allowing surface thaws. The maximum value of $T_{\text{max}}$ is 268 K for 0.24 bar and 0\% cloud coverage, while a second, local maximum equal to 265 K can be found at 5.62 bar and 100\% coverage.

At variance with the models using H$_2$O Earth-like clouds, in A0F0 there is only a minor increase in the sensitivity to $P_0$ variations when the cloud fraction changes, and the maximum variability is produced when coverage is 0\%, rather than 100\%. There is no coverage value that makes the model insensitive to pressure changes as in B0GH, BLGH and C0GN and in general, the impact of cloud fraction is smaller at any given pressure value.


\subsection{Combined effects of the obliquity, eccentricity and argument of perihelion}\label{ss:orbit_results}

\begin{figure*}
\plottwo{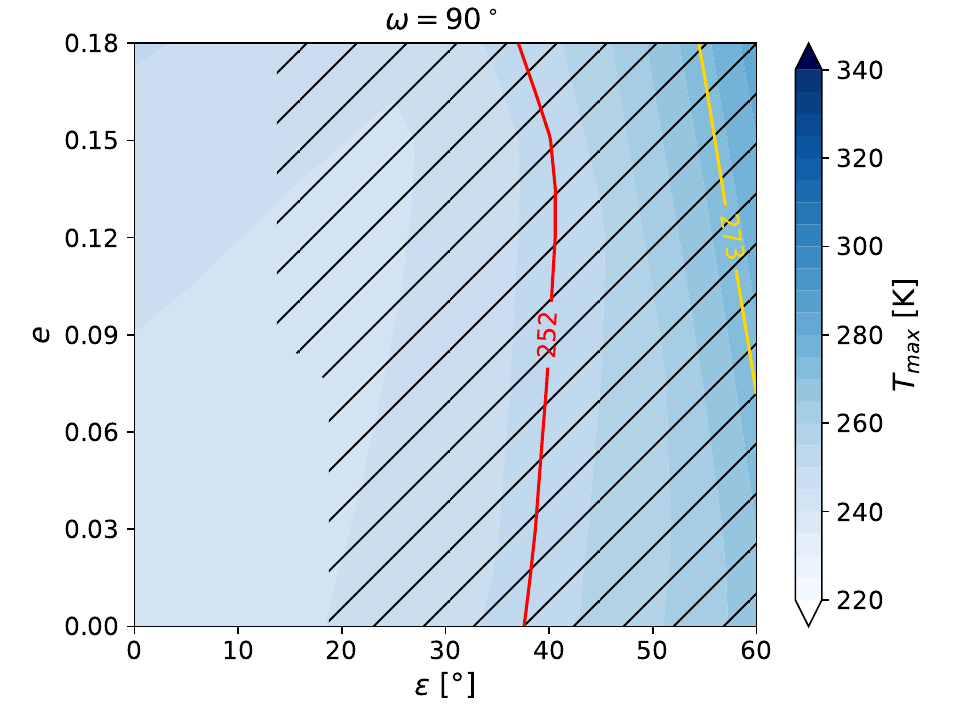}{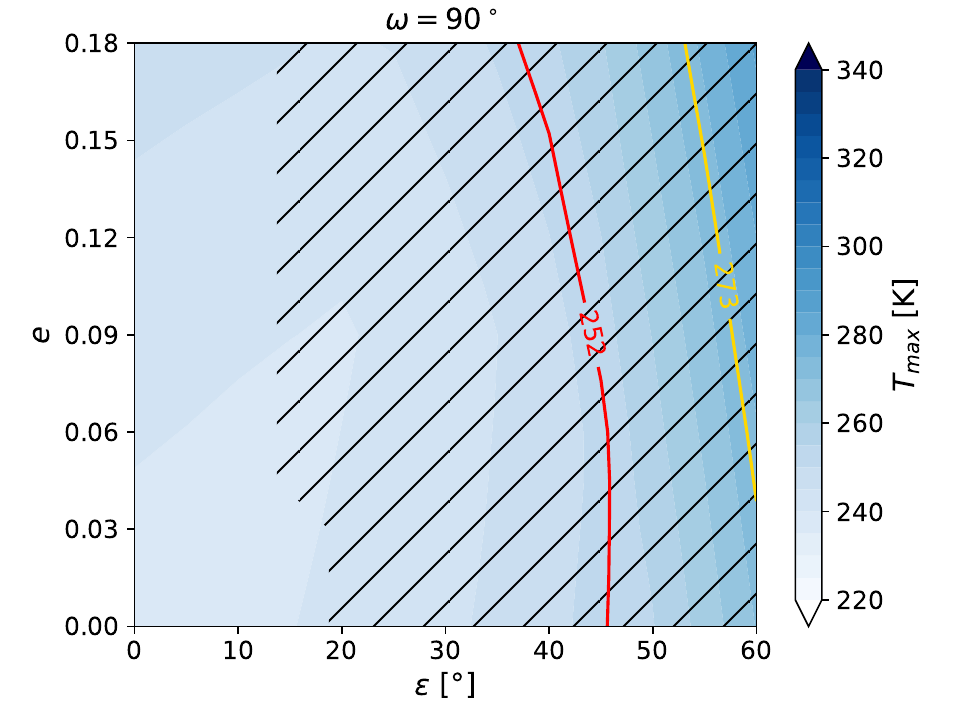}
\plottwo{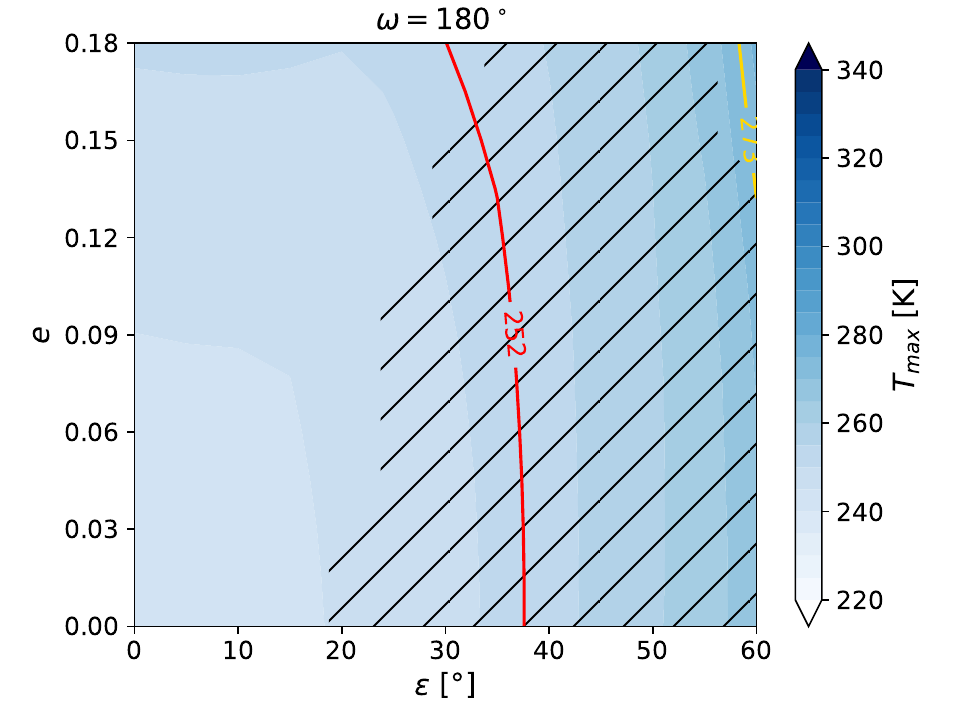}{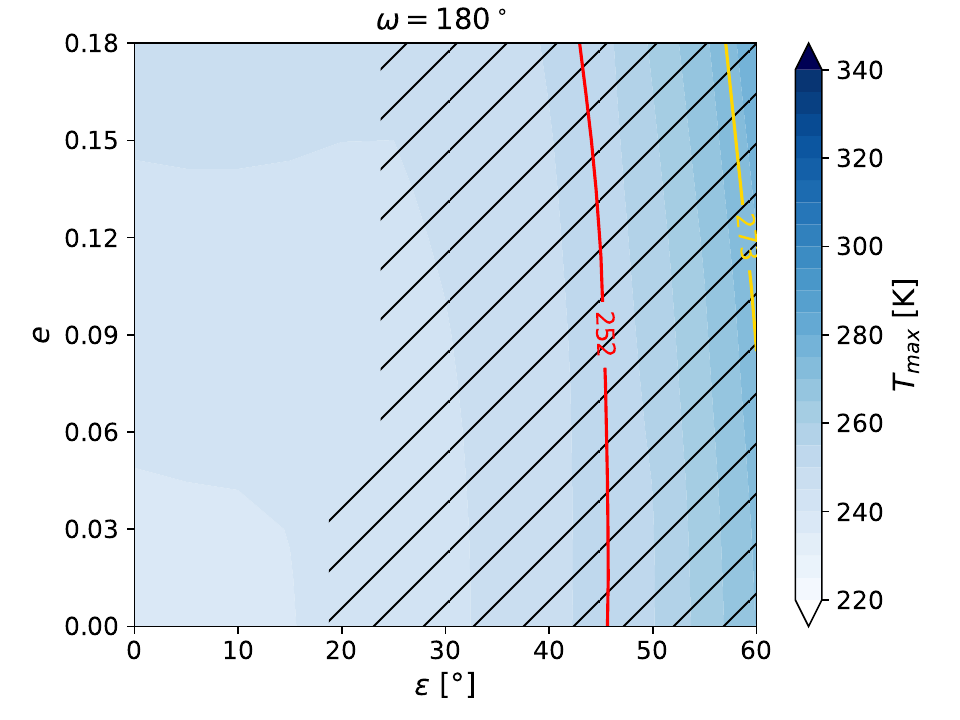}
\plottwo{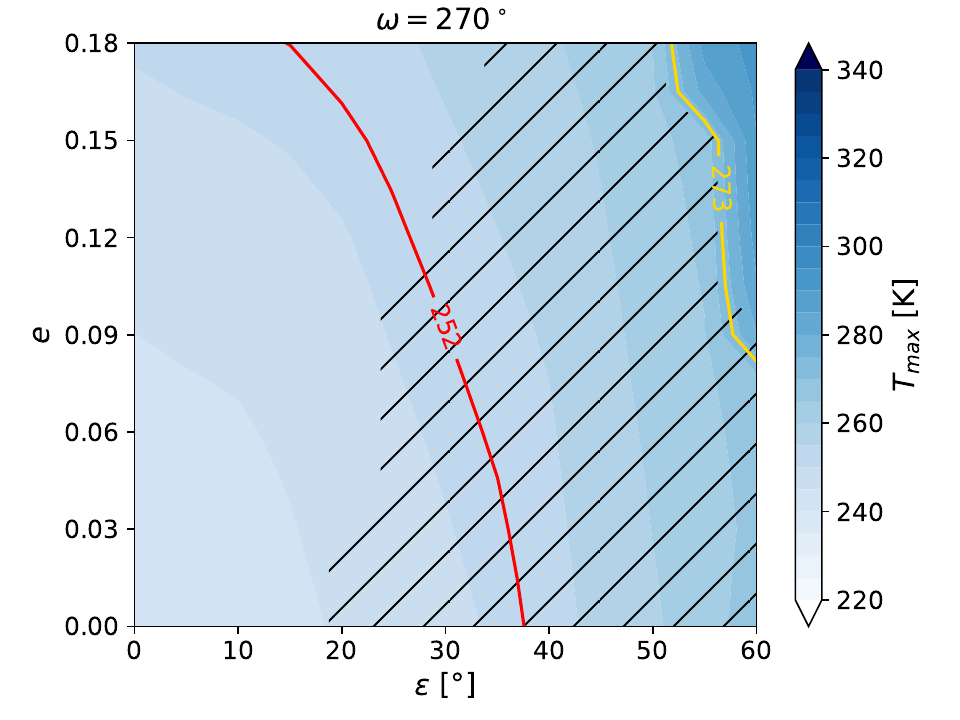}{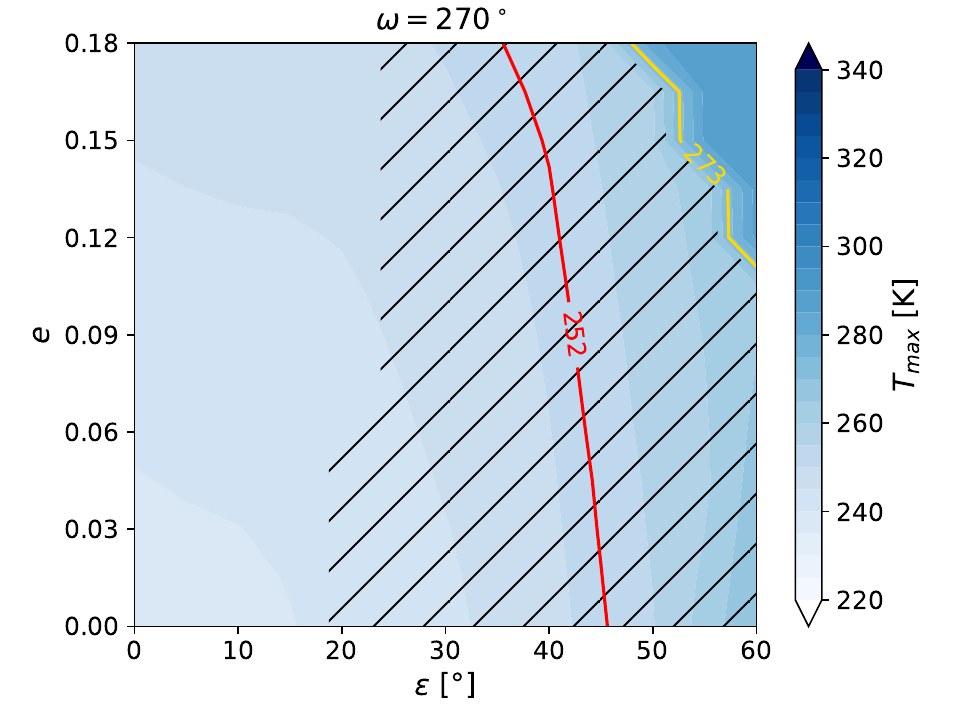}
\caption{The maximum surface temperature $T_{\text{max}}$ as a function of the eccentricity and the obliquity, for a fixed surface pressure of 1.33 bar. Left column: model B0GH. Right column: model C0GN. Top row: $\omega=90^\circ$ (perihelion at southern hemisphere's summer solstice). Middle row: $\omega=180^\circ$ (perihelion at equinox). Bottom row: $\omega=270^\circ$ (perihelion at northern hemisphere's summer solstice).  \label{fig:results_orbits}}
\end{figure*}

We can now study the combined effect of eccentricity and obliquity, for a fixed $P_0$. We chose the models B0GH and C0GN for this analysis and we varied $e$ and $\varepsilon$ in the same interval and with the same spacing adopted in Sections \ref{ss:obliquity_results} and \ref{ss:eccentricity_results}. The surface pressure has been fixed at 1.33 bar, because it is the highest $P_0$ value for which atmospheric condensation can either occur or not depending on the obliquity. In other words, it is the $P_0$ value for which the output is maximally dependent on the specific choice of $e$ and $\varepsilon$, and thus the most interesting to investigate. Each obliquity-eccentricity combination has been run three times for different values of the argument of perihelion, namely $180^\circ$, $270^\circ$ and $90^\circ$.

The results are reported in Fig.~\ref{fig:results_orbits}. First of all, this test confirms that obliquity is the main driver of seasonality, since $T_{\text{max}}$ variations are more pronounced on the horizontal, rather than on the vertical axis, except for very low ($<10^\circ$) obliquities. In general, the maximum $T_{\text{max}}$ value is found for the highest tested $e$ and $\varepsilon$ values, even though its value is different when $\omega$ changes. In particular, for model B0GH, the $T_{\text{max}}$ maxima are 276, 292 and 283 K for $\omega=180^\circ$, $270^\circ$ and $90^\circ$, respectively. For model C0GN they are 279, 288 and 286 K. For a limiting scenario as the one studied here, these differences are sufficient to completely prevent seasonal thaws.

Similarly, varying $\omega$ changes the regions of the $e-\varepsilon$ plane that allows for a non-condensing atmosphere. Condensation usually occurs at the South Pole due to the latitudinal altimetric profile of Mars. Thus, the stronger southern seasonal temperature excursions produced when $\omega=90^\circ$ make it easier to cross the condensation curve of CO$_2$ during the colder South Pole winters. The opposite is true when $\omega=270^\circ$: it is the northern hemisphere that is exposed to the strongest seasonality, but due to the combined effect of altimetry, the atmosphere is stable under a wider range of eccentricity-obliquity combinations. The shape of the condensation region underlines again the role $\omega$: in the $180^\circ$ and $270^\circ$ cases, increasing $e$ shrinks it, while in the $90^\circ$ case widens it, preventing stability at lower $\varepsilon$.

The structure of these results is similar also for the models BLGH and BHGH (not shown), but the threshold $P_0$ for a non-trivial structure of the condensation and thawing regions in the $e-\varepsilon$ plane is different. For BLGH, this value is 1.0 bar, while for BHGH is 0.75 bar and since these pressures are lower, the changes of $T_{\text{max}}$ are more pronounced. On the other hand, for $P_0=1.33$ bar, most combinations allow for stable atmospheres (again, $\omega=90^\circ$ produces a higher number of runs with CO$_2$ surface condensation than $\omega=270^\circ$) and roughly $\sim 25-35$\% of them allow for seasonal thaws.

\subsection{Duration of the thaw seasons}\label{ss:timefraction_results}

\begin{figure*}
\plottwo{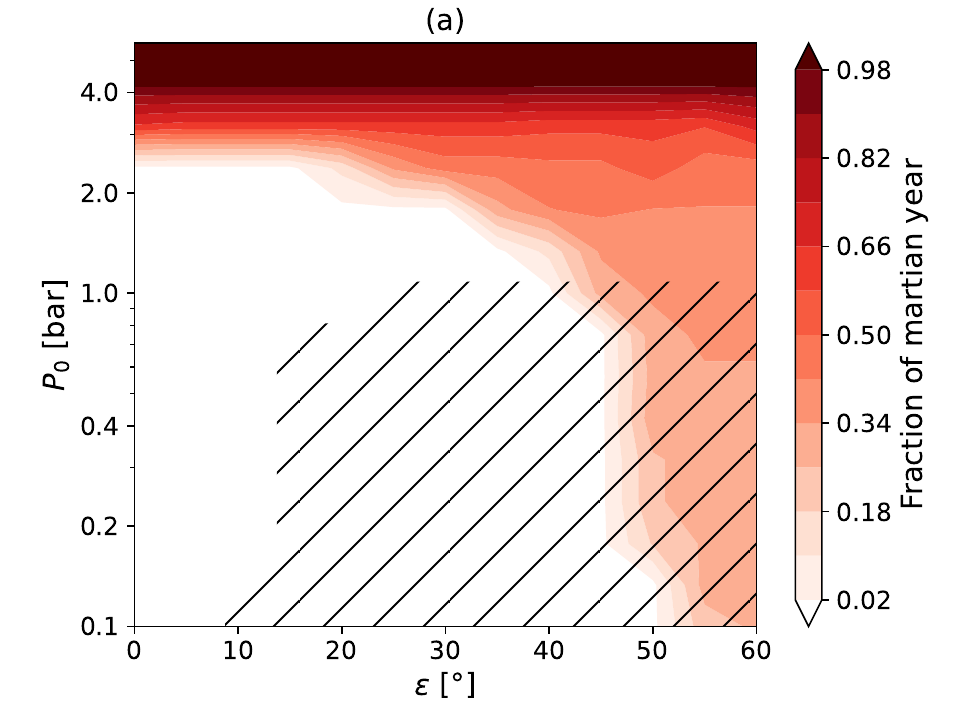}{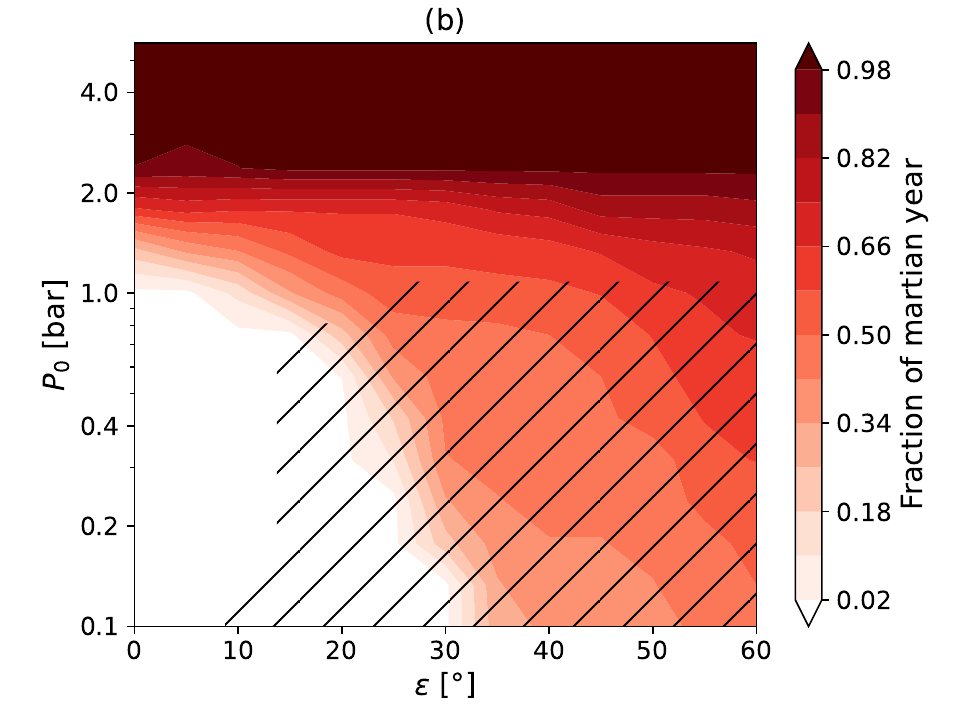}
\plottwo{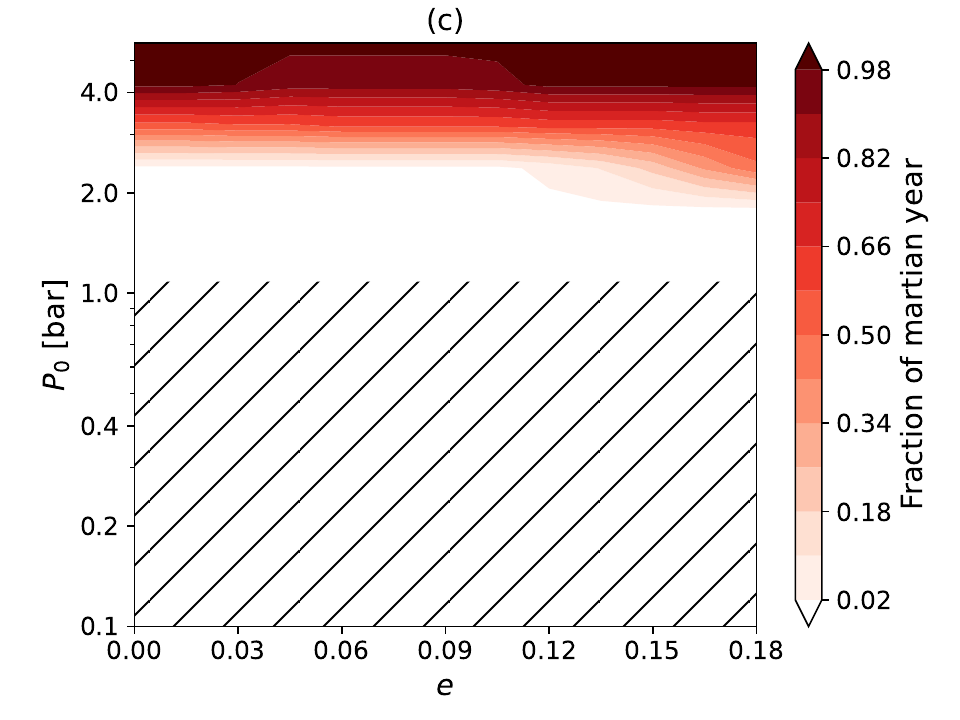}{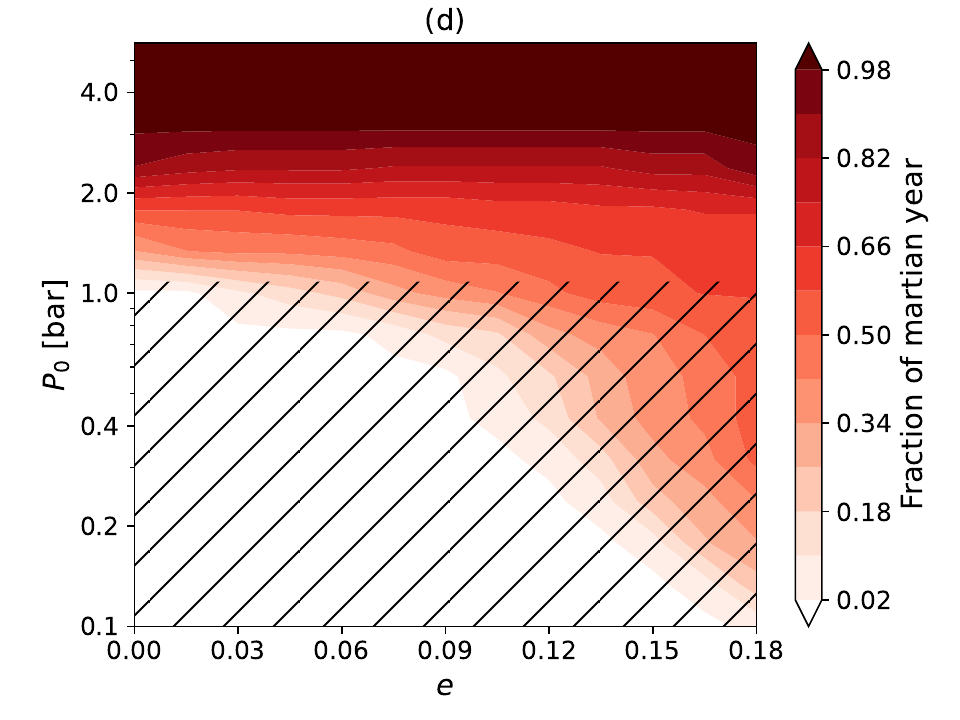}
\plottwo{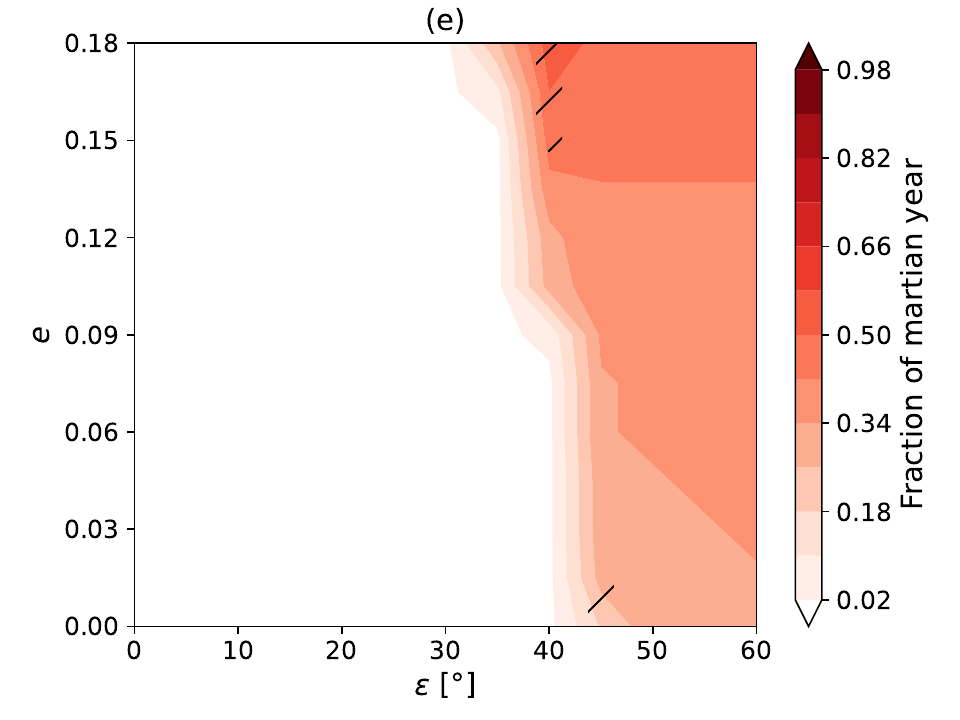}{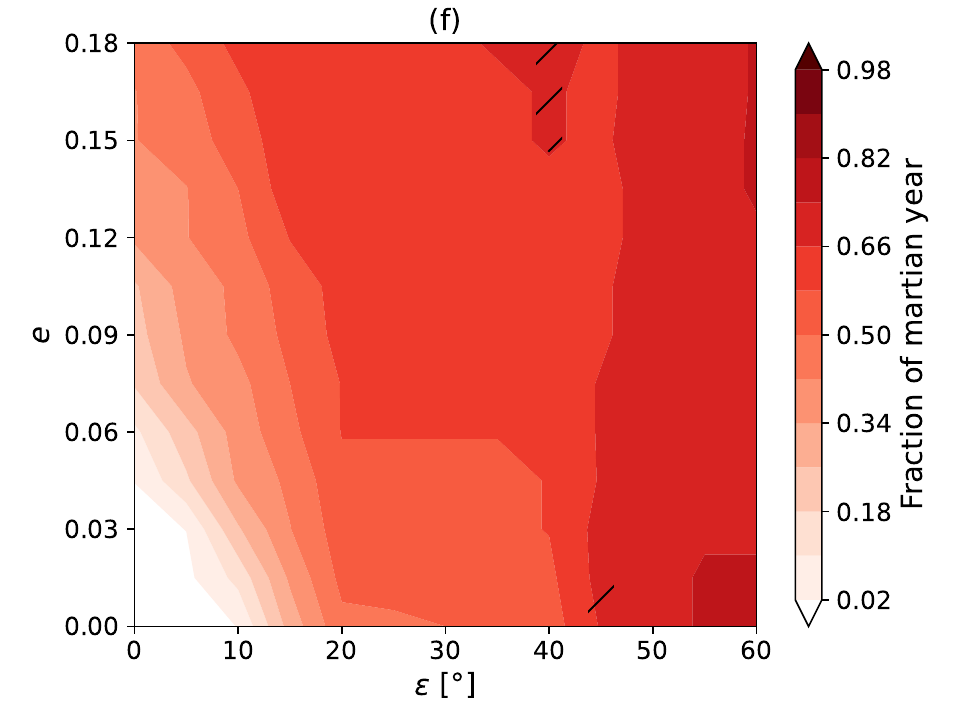}
\caption{The fraction of $P_{\text{orb}}$ during which the surface temperature remains above 273 K (left column) or above 252 K (right column) in at least one latitudinal band. All plots refer to the model BLGH. Top row: dependence on pressure and obliquity. Mid row: dependence on pressure and eccentricity. Bottom row: dependence on eccentricity and obliquity for $\omega=270^\circ$. Hatched regions refer to combinations of parameters that causes the CO$_2$ in the atmosphere to condense at surface. \label{fig:results_timefraction}}
\end{figure*}

Finally, we analyze the duration of the thaw phases. The existence, in the latitude-orbital phase plane for a given ESTM run, of point where $T_s$ is above a given threshold (e.g. 273 K) does not ensure that an actual thaw can happen. The duration of favorable conditions for deglaciation must be taken into consideration. These conditions last differently at different latitudes. In order to simplify the problem, we consider only the latitude at which $T_{\text{max}}$ occurs, since in our EBM it usually corresponds to the latitude for which the above-freezing season last the longest, and we express it in terms of fraction of martian year.

The results for the BLGH model are reported in Fig.~\ref{fig:results_timefraction}. The other models are not shown but the general considerations that we draw here apply also on them. A first distinction can be made between the region of each parameter space in which deglaciated conditions are permanent from the region in which they are seasonal. Permanently warm latitudinal bands are possible only above a certain pressure threshold, independently of the other parameter varied, and are associate with temperate global conditions (i.e. the "warm-and-wet" Early Mars scenario). Wherever the existence of a thaw season is primarily linked to the $P_0$ choice (e.g. in the pressure-eccentricity plane), then the seasonality region is small: either the parameter set always prevent deglaciation, or they always allow it. This is not strange considering that, if thaws are driven by the presence of relatively high pressures, then the more efficient heat distribution limits the local seasonal variations of temperature.

On the other hand, seasonal thaws are possible at all pressures, even though in some cases are prevented by atmospheric collapse (see e.g. panel a in Fig.~\ref{fig:results_timefraction}). The duration of the warm season can in principle be very short, but if it is present, it is usually longer than $\sim0.15$ of martian year ($\gtrsim 100$ d). In other words, when thawing can occur, then it is not limited to negligibly short ($\lesssim 0.1$) fractions\footnote{Since we calculate the $T_s$ for 48 points along the martian orbit, the minimum possible fraction would be $\sim 0.02$.}. For example, a typical fraction value along the border between the non-deglaciating and the deglaciating regions for a 273 K threshold temperature (Fig.~\ref{fig:results_timefraction}, panels a, c and e) is 0.3. This happens because $T_{\text{max}}$ nearly always occur in the northern hemisphere, where the surface is covered in water: when ice melts the albedo decreases, making it harder for ice to form again. Thus, this is basically a manifestation of a bistability mechanism, albeit on a regional (rather than a global) scale.

\section{Discussion} \label{sec:discussion}

\begin{figure*}
\plotone{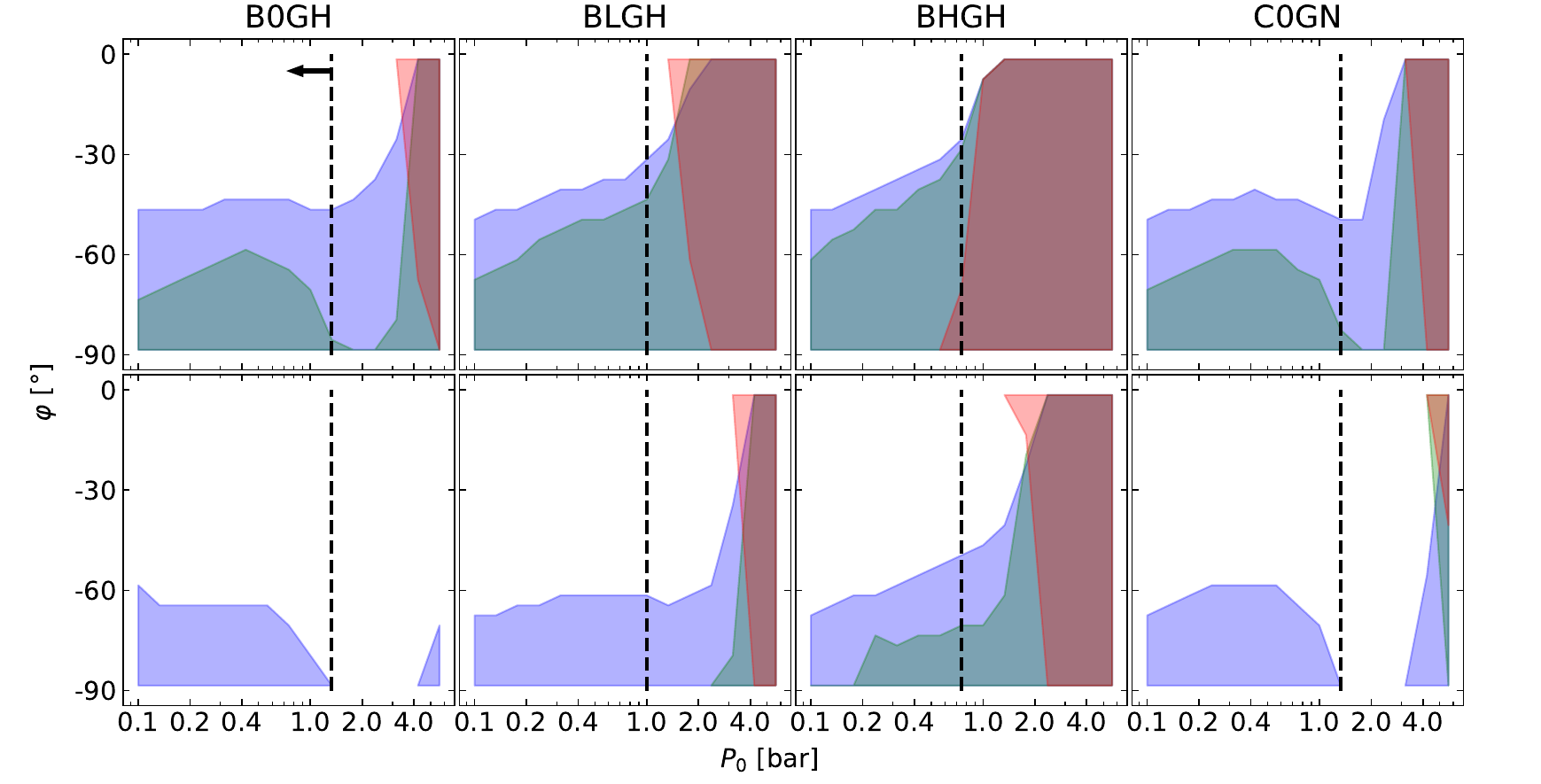}
\caption{The latitude interval in which the temperature at $h=0$ m is higher than $T_{\text{thr}}$ for at least one time step, as a function of the datum pressure $P_0$. Results refer to the martian southern hemisphere. Top row: $T_{\text{thr}}$ = 252 K. Bottom row: $T_{\text{thr}}$ = 273 K. From left to right: B0GH, BLGH, BHGH and C0GN models. Red, green and blue shaded areas refer to cases with $\varepsilon$ = $30^\circ$, $45^\circ$ and $60^\circ$, respectively. The vertical dashed lines mark the pressure below which surface atmospheric condensation occurs for each of the shown combinations of pressure and $\varepsilon$. \label{fig:discussion_valleys}}
\end{figure*}

Searching for seasonally, rather than continuously, temperate conditions on Early Mars relaxes the constraints on the pressure and the general composition of the atmosphere. This is especially true when the orbital parameters of the planet are substantially different from their modern values and might help to reconcile the somewhat paradoxical observational evidences inferred from martian geological records. This desirable outcome is especially evident in the results presented in Sections \ref{ss:obliquity_results}, \ref{ss:orbit_results} and \ref{ss:timefraction_results}.

From Fig.~\ref{fig:results_obliquity}, it appears that additional greenhouse gases (such as H$_2$ or CH$_4$) are not strictly needed, since when obliquity is sufficiently high seasonal thaws can happen even if the atmosphere is made only of CO$_2$ with a minor contribution from H$_2$O (using desert-like RH values). Even minor greenhouse contributions from other gases make the parameter space region in which seasonal liquid water conditions are possible substantially larger. As an example, the concentrations of H$_2$ needed to produce a stable, year-long warm-and-wet Mars seems difficult to produce using reasonable outgassing and atmospheric escape rates \citep{ramirez14}, but considering the seasonal thermal maxima, as we did here, would substantially lower such a requirement.

Fig.~\ref{fig:results_orbits} hints to a possible mechanism, different from adding greenhouse gases, to locally warm up  Early Mars, which is varying together the eccentricity, the obliquity and the argument of perihelion. While the combinations of these parameters that allow for thaws are unlikely (at least for an atmosphere without H$_2$ or CH$_4$), their existence suggest a possible pathway for the valley networks formation that is not as explored as others. 

Fig.~\ref{fig:results_timefraction} shows instead the different pressure range required for a fully vs a seasonally deglaciated planet. Depending on the exact value of other parameters, seasonal deglaciation requires at the very least 25\% less surface pressure for a given atmospheric composition, in some cases allowing long thaws even at 0.1 bar (panels a and b). The biggest problem in this regard seems to be the condensation of the atmosphere at surface. Specific recipes would be required to track the climate evolution under this kind of conditions. Upper limits on martian paleopressure are uncertain, but generally point to a relatively thin atmosphere. \cite{kite14} estimates a value of about 2 bar from the crater diameter distribution at the beginning of the Hesperian age (3.6 Gyr ago), while \cite{edwards15}, by studying known martian carbonate deposits, found that it is improbable that more than $\sim$0.5 bar has been sequestered at surface. By combining their results with the upper limit on CO$_2$ escape \citep[0.15 bar,][]{lammer13}, a reasonable maximum pressure of $\sim$0.7 bar during the Hesperian can be obtained. Interestingly, our model BHGH is both stable and able to produce seasonal thaws at around that pressure value, albeit for a narrow ranges of obliquity and eccentricity values. On the other hands, several combinations of atmospheric and planetary parameters among those tested here are able to produce seasonal liquid water conditions in the 1-to-2 bar range.

\subsection{Comparison with the observed distribution of valleys} \label{ss:comparison_observations}

We can now compare our results with the observed latitudinal distribution of valley networks on the martian surface. \cite{hynek10} reported that a majority of the Noachian valley networks (which represents 91\% of the total valley networks identified by them) can be found between 45$^\circ$S and $5^\circ$N, with a peak a few degrees south of the equator, while Hesperian networks present a more uniform distribution with a first peak at $35^\circ$S and a second one at $5^\circ$N. The altitude distribution of Noachian network midpoints is nearly gaussian and peaks at 1500 m above the datum (roughly corresponding to the mean altitude of Noachian terrains), while Hesperian networks are again distributed more regularly with a mean altitude consistent with 0 m.

Due to the effect of the lapse rate, the latitude at which $T_{\text{max}}$ occurs is nearly always in the northern hemisphere. However, Noachian and Hesperian valleys are far more often found in the South. Thus, we can check if sufficiently warm conditions can occur also in the latitudinal regions associated with the presence of these valleys. This is especially relevant since liquid water conditions tend to occur at high obliquity values: in these cases, higher latitudes are subject to more intense $T_s$ variations than the equator and thus deglaciations are favored at high, rather than low, latitudes.

As it is possible to see in Fig.~\ref{fig:discussion_valleys}, models B0GH and C0GN struggle to produce conditions compatible with liquid water runoffs at low southern latitudes, even for the lower $T_{\text{thr}}$ considered. On the other hand, models BLGH and BHGH allow brine to stay liquid for at least the summer season across the entire southern hemisphere when $P_0$ is above 1.77 and 1.33 bar, respectively, while they require $\gtrsim 2$ bar of pressure to let pure water ice to melt in the $0^\circ$S - $45^\circ$S region.

Concerning the dependences on parameters other than the obliquity (not shown in the Figure), the impact of eccentricity and  surface albedo is smaller, while that of the cloud coverage is larger, thus confirming the trends encountered in Section \ref{sec:results}. For model BHGH, thaws are possible across most of the southern hemisphere for $P_0>1$ bar if we consider brine, and $P_0>2$ bar if we consider water independently of the chosen eccentricity. Pressure limits move to 2 bar for brine and 3 bar for water in the BLGH case. C0GN requires more than 4 bar to allow for water runoffs in the south, while B0GH never produces datum temperatures at or above 273 K. Land albedo also has a minor impact on southern summer temperatures, despite the fact that oceans are mostly absent in that region. Again, the main determinant are the surface pressure and the composition. For example, in model BHGH, lowering the albedo from 0.41 to 0.11 only lowers the required pressure from 2.0 to 1.33 bar (from 3 to 2 bar for BLGH). On the other hand, the cloud fraction has a larger impact due to their important greenhouse contribution in our model. Cloud cover fraction at or above 0.5 are capable of warming the southern hemisphere to a sufficient level to allow thaws in all models, albeit above different $P_0$ threshold values.

At variance with Hesperian valley networks, Noachian ones lie at 1500 m of altitude, as previously noted. This places a tighter constraint on the climate state of the planet, since at that height the surface temperature is 7-8 K lower than at the datum. We thus repeated the analysis including this effect, finding that it increases the pressure threshold for seasonal thaws by $\sim$33\%. Again, the BHGH model is the best suited to explain the observational evidences, while B0GH and C0GN can produce brine runoffs but not water ones.

An interesting aspect of searching for seasonal thaws is their possible role in the formation process of Equatorial Layered Deposits \citep[ELDs,][]{schmidt21}. ELDs are stratified deposits mostly found in craters and basins in the equatorial region (especially in Terra Arabia) and probably associated with water activity. A variety of processes have been proposed as their source \citep[see e.g.][Table 1 and 2]{schmidt22b}. The thinning-thickening sequence of the deposits might be explained either by a relatively strong seasonal variability (Schmidt, G., private communication) or by secular changes in the hydrological cycle forced by obliquity and eccentricity variations, depending on the time scale associated with this feature and that is currently under investigation. Morphological evidences found in the Gale Crater evaporitic basin seems to point toward a sustained wet-dry cycling compatible with sustained seasonal variations \citep{rapin23} as those found in some of our simulations.

\section{Conclusions and future prospects} \label{sec:conclusions}

In this work we have studied the combined impact of varying the surface pressure and other martian planetary parameters (obliquity, eccentricity, argument of perihelion, surface albedo and fraction of clouds) on the seasonal temperature changes in order to identify the conditions conductive to regional ice melting. We tested five different atmospheric compositions with varying H$_2$O ($RH=$ 0\%, 20\% and 40\%) and CH$_4$ (0\%, 0.1\% and 1\%). We employed a seasonal-latitudinal climate model (ESTM) coupled with an up-to-date RT code (EOS), including the effects of the band-averaged altimetry, a northern ocean with either 150 or 550 m of GEL and checking for possible CO$_2$ condensation at surface. In total, we run $\sim$10$^4$ cases (i.e. different combinations of input parameters).

Our main findings can be summarized as follows:
\begin{enumerate}
    \item If both the planet obliquity and datum pressure are sufficiently high ($\varepsilon \ge 50^\circ$ and $P_0 \ge 2.0$ bar) then no other additional greenhouse gases other than CO$_2$ and desert-like levels of H$_2$O are needed to produce seasonal thaws in the northern hemisphere. These conditions also guarantee that no CO$_2$ condensation happens at the surface. However, the same models fail to produce seasonal melting in the southern highlands.
    \item Adding a 0.1\% of CH$_4$ lowers the requirements for northern thaws to 1.33 bar (at $\varepsilon=45^\circ$) and simultaneously allows for melting in the $60^\circ$S - $90^\circ$S region. Increasing the pressure to 3.5 bars or above melts also the region near the equator. If $f_{\text{CH4}}$ is instead set to 1\%, both northern and southern thaws are possible independently of the obliquity when $P_0 \ge 1.33$ bar.
    \item Increasing the eccentricity allows for thaws and stabilizes the atmosphere when combined with $\varepsilon \ge 50^\circ$ and $\omega \sim 270^\circ$ (i.e. when Mars is at perihelion during the northern hemisphere's summer solstice). The effect of eccentricity and obliquity alone can change $T_{\text{max}}$ by 29-52 K, depending on the specific choice of the argument of perihelion. This effect is important if the planet is near the threshold conditions for hosting liquid water.
    \item Changing the land albedo has some impact in the models without CH$_4$, while it is less important when CH$_4$ is present. If $f_{\text{CH4}}$ is set to 1\%, both northern and southern melting are possible at $P_0 \ge 2$ bars independently of the land albedo. This is especially interesting when we consider that the southern hemisphere has a very low ocean fraction.
    \item Clouds are difficult to treat and have been included in this work by parameterizing their contributions to OLR and TOA albedo. Clouds calibrated on the \cite{forget13} model (i.e. CO$_2$-ice clouds) give only a modest contribution in terms of both average temperature and $T_{\text{max}}$ when coupled with a dry CO$_2$-dominated atmosphere. However, they can prevent the atmospheric collapse if fractional coverage is higher than 0.8. Clouds calibrated on the modern Earth model (i.e. H$_2$O clouds) are more efficient at warming Mars. We found that the minimum cloud coverage to prevent the atmospheric collapse is 0.5 if there is no CH$_4$, 0.2 when $f_{\text{CH4}}$ is at 0.1\% and are not needed when $f_{\text{CH4}}$ is at 1\%.
    \item If melting is possible, then mild conditions last usually more than 15\% of martian year. While we did not calculate the amount of liquid water that we can obtain, we are confident that it is not negligible (as if the thawing season were very short).
    \item Changing the GEL level has a relatively small impact on the results. Notably, the transition from the region of the parameter space in which seasonal thaws are not allowed to that in which they can happen is sharper. This is due to the fact that a larger ocean causes a larger swing in the average surface albedo when it melts.
    \item If we relax the thawing conditions adopting the freezing temperature of brine instead of that of pure water, the combinations of planetary and atmospheric parameters that allow melting significantly widen. Seasonal brine runoffs in the northern hemisphere are allowed also in dry CO$_2$-dominated atmospheres with high ($\ge 0.8$) CO$_2$-ice cloud fractions for pressures above 1.33 bars.
\end{enumerate}

To conclude, we confirm the previous results indicating the necessity of other greenhouse gases other than CO$_2$ and H$_2$O to allow for an hydrological cycle that involves the southern highlands. However, we show that depending on the combinations of orbital ($\varepsilon$, $e$ and $\omega$) and planetary (land albedo, cloud coverage) parameters, this contribution can be significantly lower than usually expected \citep[$\sim$2-3\% of H$_2$ in a 2 bar CO$_2$ atmosphere,][]{ramirez17b,wordsworth17} and probably easier to explain within the other known geological constraints of Mars. In particular, to our knowledge, no previous investigations on the interplay between eccentricity, obliquity and the argument of perihelion for different atmospheric models were performed. An investigation on the interplay between eccentricity, obliquity and the argument of perihelion has been recently performed to explain the formation of gullies in the recent past of Mars \citep{dickson23}, but here we explored a larger portion of the parameter space and referred to Early Mars conditions. We also underline the advantages of using seasonal-latitudinal, rather than single-column, climate models to combine together computational efficiency and predictive power. 

There are several possible avenues for future developments. First of all, we limited ourselves to a very narrow range of atmospheric chemical compositions, all of which in equilibrium. Using the same methodology, it would be possible to investigate the effects of pulses of e.g. volcanic gases for different combinations of martian orbital and surface parameters. Second, we took into consideration a simple parameterization of clouds, rather than including them into the RT model. Improving the RT modeling of clouds is very important, since as we have shown, their contribution to planetary warming is crucial when the concentrations of other greenhouse gases is low. Third, it would be desirable to calculate the amount of liquid water that is made available during the seasonal thaws. A precise assessment of precipitation rates requires 3D GCMs, while an upgraded version of ESTM would be able to provide approximate latitudinally-averaged estimates that can lend additional robustness to these findings.

\begin{acknowledgments}

P.S. wants to thank Gene Walter Schmidt for the insightful discussion concerning the ELDs and Ramses Ramirez for the useful suggestions. The Authors also thanks the anonymous referee for the careful reading of the manuscript and the valuable comments. We acknowledge support by the Italian Space Agency with the \textit{Life in Space} (ASI N. 2019-3-U.0) and \textit{ASTERIA} (ASI N. 2023-5-U.0) projects, OGS and CINECA with the HPC-TRES program award N. 2022-02 and INAF with the mini-grant titled \textit{Radiative models for the paleoatmospheres of
Mars, Earth and Venus} (F.O. 1.05.121.04.03).

\end{acknowledgments}

%

\vspace{5mm}





\appendix

\section{Lapse rate calculation} \label{app:eos}

\begin{figure*}
\plottwo{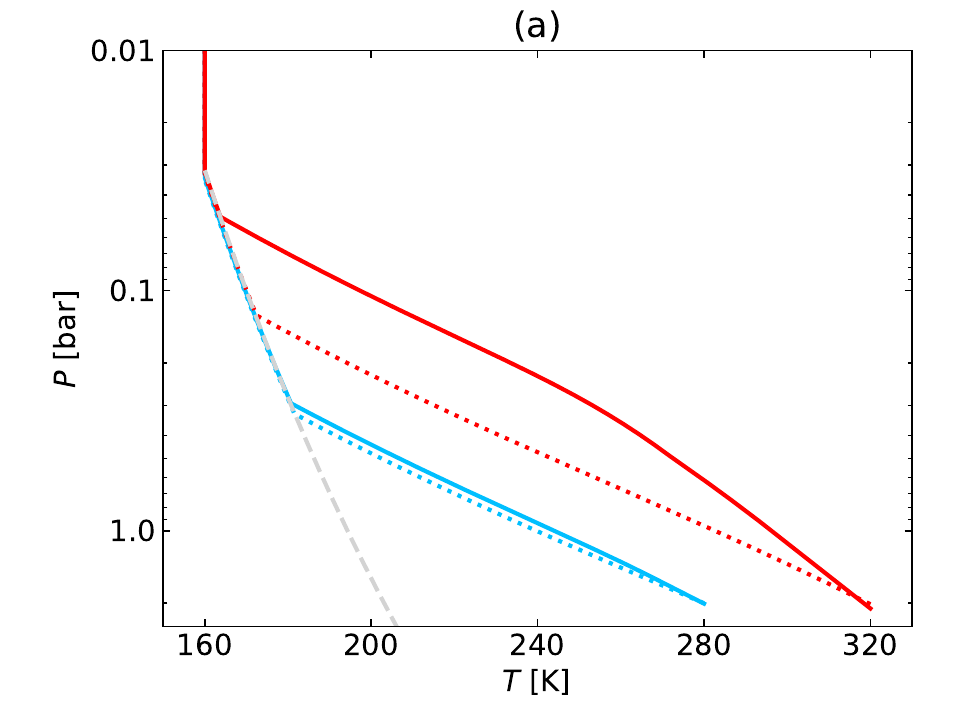}{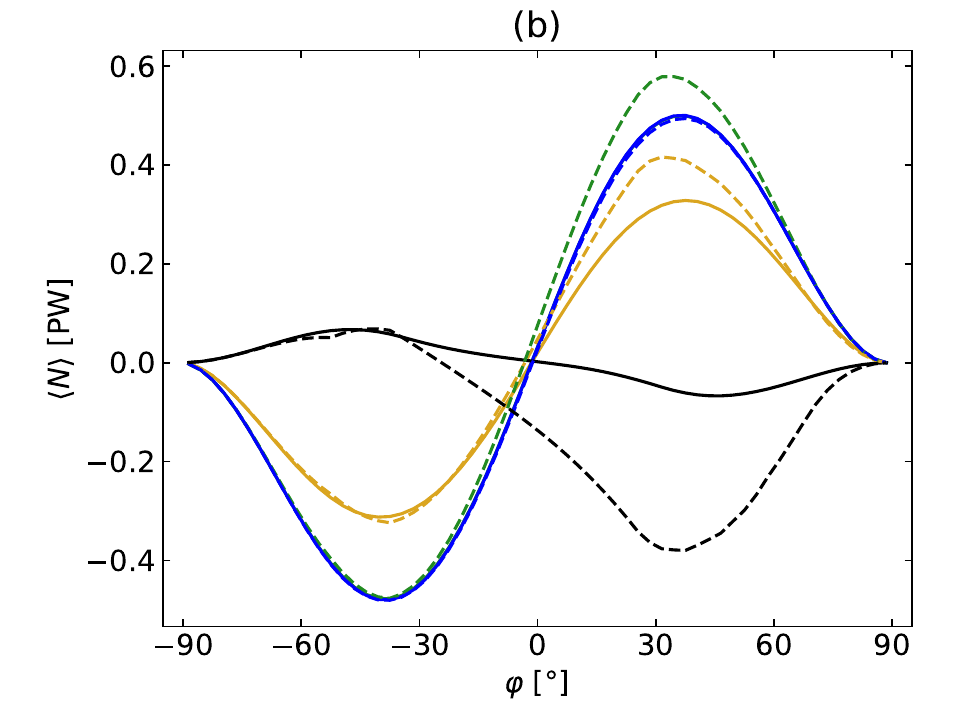}
\plottwo{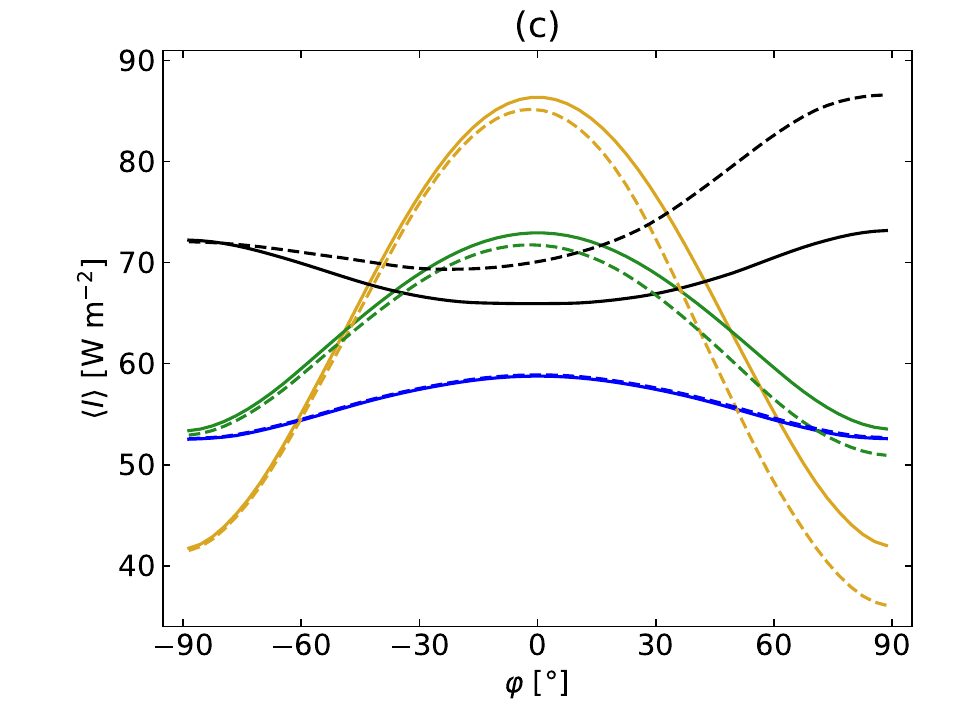}{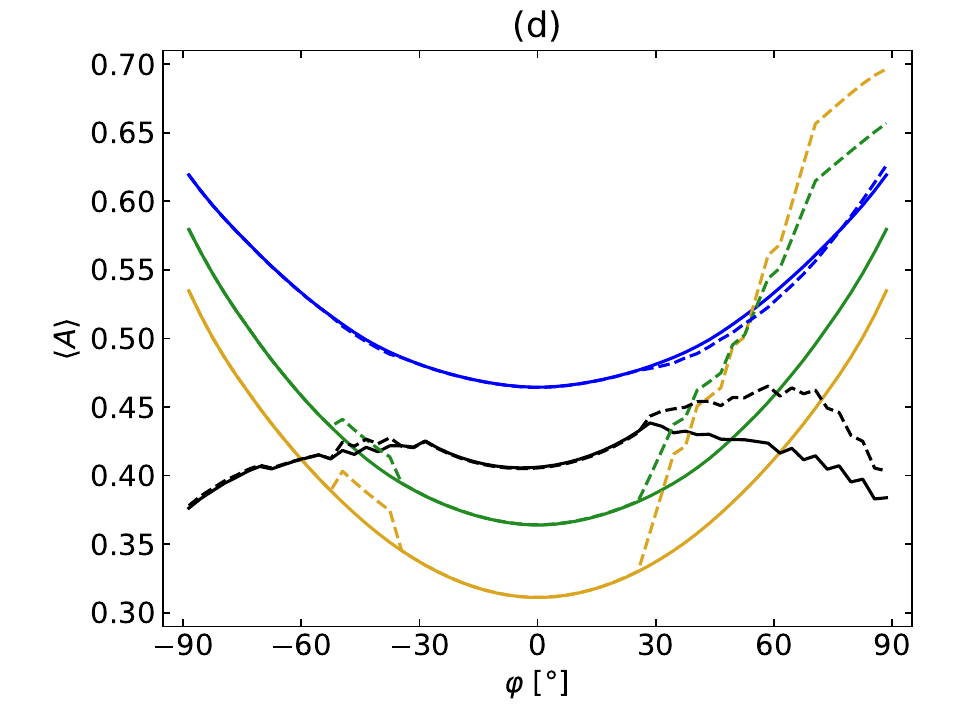}
\caption{Panel (a): Four examples of vertical PT profiles calculated using the procedure described in this Section. Cyan and red lines are calculated specifying a $T_s$ equal to 280 K and 320 K, respectively. Solid refers to cases with $RH=100\%$ and dotted to cases with $RH=20\%$. The grey dashed line is the condensation curve for CO$_2$ from the \cite{kasting91} formulation. Panel (b): the yearly averaged net heat flux as a function of latitude for eight cases studied in the paper. Yellow, green and blue lines describe, respectively, to cases with $P_0 =$ 0.1, 1.0 and 5.62 bar and $\varepsilon = 0^\circ$. Black line describes to the case with $P_0 = 1.0$ bar and $\varepsilon = 60^\circ$. Solid and dashed lines refer, respectively, to the BLH0 and BLGH models. Solid green and blue lines are nearly ovelapped. Panel (c): the yearly averaged OLR $\langle I \rangle$ for the same cases as in (b). Panel (d): the yearly averaged TOA albedo $\langle A \rangle$ for the same cases as in (b). \label{fig:appendix_ptprofile}}
\end{figure*}

Both the vertical pressure-temperature profile and the altimetric adjustment have been calculated using a modified version of the moist pseudo-adiabatic lapse rate $\Gamma$ provided by the American Meteorological Society Glossary\footnote{\url{https://glossary.ametsoc.org/wiki/}}:
\begin{equation}
\Gamma = g \, \frac{(1 + r_v)(1 + \frac{L_v r_v}{R_d T})}{c_{pd} + r_v c_{pv} + \frac{L_v^2 r_v (\phi + r_v)}{R_dT^2}}
\label{eq:lapserate}
\end{equation}
where $g$ is the gravitational acceleration, $r_v$ is the mass mixing ratio of the condensible, $L_v$ is the latent heat of vaporization of the condensible, $R_d$ is the gas constant of the dry gas, $c_{pd}$ and $c_{pv}$ are the specific isobaric heats of the dry gas and the condensible, respectively, and $\phi$ is the ratio between the molecular masses of the condensible and the dry gas. Dry gas properties are the weighted average of the component gases considered (here, CO$_2$ and CH$_4$). From a theoretical point of view, this lapse rate is valid only if the $RH$ of the condensible is 100\%, while the dry lapse rate should be used if this condition is not true. However, we do know that the real lapse rate in the Earth atmosphere is generally in-between the moist and the dry lapse rate due to inhomogeneities in the convecting air mass. To represent this fact, the $r_v$ term in the H$_2$O condensing part of the troposphere is scaled by the $RH$ value specified in Tab.~\ref{tab:models}.

The condensible is H$_2$O in the lower troposphere and CO$_2$ in the upper troposphere. In practice, we integrate Eq.~\ref{eq:lapserate} on a 100 m-spaced grid in height (10 m-spaced grid when we calculate the altimetric correction) starting from the surface and up to the height corresponding to a specified pressure (i.e. the $P_{\text{TOA}}$) or temperature (i.e. the stratospheric temperature). The terms $r_v$, $L_v$, $c_{pd}$ and $c_{pv}$ depend on temperature and are updated accordingly by iterating on Eq.~\ref{eq:lapserate}. At each step we check if CO$_2$ condensation occurs and if this is the case, we begin treating CO$_2$ as the condensible in place of water. Strictly speaking H$_2$O condensation continue to happen, but since it contributes a negligible amount of latent heat with respect to CO$_2$ in this region, we neglect it. The altitude-temperature profile obtained is then converted into a pressure-temperature profile using the barometric formula for a non-isothermal atmosphere:
\begin{equation}
P_{i+1} = P_i\,\biggl(\frac{T_{i+1}}{T_i}\biggl)^{-g/(R_{m}\Gamma)}
\end{equation}
where $T_i$ is the temperature of the $i$th layer and $R_m$ is the gas constant of the atmospheric mix (dry plus moist). Fig.~\ref{fig:appendix_ptprofile}, panel (a), shows four examples of PT profiles calculated using this procedure. There is a small offset with respect to the CO$_2$ condensation curve due to the fact that these profiles refer to the case with $f_{\text{CH4}} = 1\%$.

Using temperature-dependent latent and isobaric specific heats allows us to capture the main contributions to deviations from an ideal gas lapse rate. In \cite{simonetti22} we adopted the formulation of \cite{kasting91}, that includes also the effects of a varying CO$_2$ compressibility. However, that formulation approximate the pressure of the condensible to zero, which is a reasonable assumption only for relatively cold ($T_s \le 280$ K) atmospheres but breaks down at higher temperatures. Low total surface pressures make the problem worse, and in general the error caused by not considering the changing gas compressibility are larger than the error caused by neglecting the vapor pressure, at least in the pressure and temperature ranges that we worked with. 

\section{Diffusion, OLR and albedo profiles of selected cases} \label{app:profiles}


As discussed in the paper, seasonal thaws can occur on the surface of Early Mars because of the imperfect heat redistribution between latitudinal bands. Since the diffusion parameter $D$ in Eq.~\ref{eq:diffusion} has a complex dependence on both input parameters and simulation variables, a more relevant quantity to show is the net flux of heat $N$ along the latitudinal direction. We show the yearly averaged value of $N$, $\langle N \rangle$, for a very small number of selected cases in Fig.~\ref{fig:appendix_ptprofile}, panel (b). A positive (negative) $N$ means a poleward flux in the northern (southern) hemisphere. We remind that, in all cases, diffusion is calculated before applying the altimetric correction or, in other words, under the hypothesis that it is dominated by processes happening at the same altitude throughout the entire planet. We start considering the cases with $\varepsilon = 0^\circ$, represented by yellow, green and blue curves, for which $N$ is nearly time-independent (apart for the effect of obliquity) and thus $\langle N \rangle \sim N$. Solid curves refer to a configuration without oceans (the BLG0 model) and, as such, are symmetric with respect to the equator. In this configuration the role of $P_0$ can be isolated: it increases by $\sim 50-60$\% at all latitudes $P_0$ is increased from 0.1 to 1 bar, while it seems constant when $P_0$ increases again to 5.62 bar. Analyzing the dependence of the peak transport $\langle N \rangle_{\text{max}}$ on $P_0$ (not shown) we notice that it reaches a peak at 2.37 bar and then slightly decreases. This is most probably due to the weakening of the dependence between the albedo and $z$ at high pressures, which reduces the difference in the absorbed radiation between the equator and poles. The addition of the ocean (dashed curves) breaks the north-south symmetry and has the effect of enhancing (when glaciated, yellow and green) or suppressing (when ice-free, blue) the latitudinal heat transfer in the northern hemisphere. If we instead consider the cases with $\varepsilon = 60^\circ$ (the black curves), we find quite a different picture. First of all, $\langle N \rangle$ has the opposite sign in both hemispheres, which means that the heat flux is equatorward for at least part of the year (i.e. during summer), and the overall effect of this inverted circulation is stronger when averaged on the entire orbit. Checking the yearly average temperatures $\langle T_0 \rangle$ (not shown) we find that polar $\langle T_0 \rangle$ are slightly higher than the equatorial one (239.4 K vs 233.7 K in the BLG0 model). The higher thermal capacity of oceans with respect to the martian surface makes the northern hemisphere a strong contributor of heat to the rest of the planet.

We complete the picture by discussing also the OLR and the TOA albedo profiles for the same cases described above. In Fig.~\ref{fig:appendix_ptprofile}, panel (c), we show the yearly averaged value of the OLR, $\langle I \rangle$. Starting again from cases with $\varepsilon = 0^\circ$, we underline the flattening of the $\langle I \rangle$ curve when pressure increases, which corresponds to a flattening of the $\langle T_0 \rangle$ curve. The presence of an ocean breaks the north-south symmetry, in particular lowering the northern longwave emission when it is glaciated (in the 0.1 and 1 bar cases) or slightly increasing it (in the 5.62 bar case). The cases with $\varepsilon = 60^\circ$ show instead an inverted curve, with higher average emissions at the poles, especially when an ocean is present.

Finally, in Fig.~\ref{fig:appendix_ptprofile}, panel (d), we show the yearly averaged value of the TOA albedo, $\langle A \rangle$. In cases with $\varepsilon = 0^\circ$, the curve is always bowl-shaped due to the fact that, at high latitudes, the $z$ angle is larger, light path is longer and thus Rayleigh scattering is more efficient. The surface albedo is also larger due to the slanted insolation. The presence of ice in models with oceans is underlined by the increase of $\langle A \rangle$ in the northern hemisphere and at the latitudes corresponding to the Hellas basin (cases with $P_0 =$ 0.1 and 1.0 bar), while the darker deglaciated ocean in the $P_0 = 5.62$ bar slightly decreases the albedo in the same bands. The cases with $\varepsilon = 0^\circ$ have a substantially different behavior. When an ocean is not present, $\langle A \rangle$ reaches a maximum at $\varphi = 30^\circ$ and then decreases. This is due to the fact that $A$ has an effect on the climate (and thus it is calculated) only when a given latitudinal band is illuminated, and for non-zero obliquities there is always a period of ``polar night'' in each hemisphere during which the Sun remains below the horizon the entire day. This means that $\langle A \rangle$ is skewed towards summertime $A$ values, which are smaller due to the lower average $z$. The irregular behavior of the curve near the poles is due to the discrete nature of the calculation. Again, the presence of an ocean increases the albedo in the northern hemisphere, which is somewhat unexpected considering that the curves in panels (b) and (c) seem to suggest it is mostly deglaciated. The apparent discrepancy comes again from the fact that $\langle A \rangle$ is calculated only considering the part of the year in which the poles are illuminated by the Sun: due to the high albedo of ices, the northern ocean remains glaciated for $\sim$60\% of the illuminated season, which increases $\langle A \rangle$. Once the ice is melted, the ocean can efficiently accumulate heat due to the low albedo, which is then released during the polar night season.

\bibliography{biblio}{}
\bibliographystyle{aasjournal}



\end{document}